\documentclass[aps,prl,twocolumn,showpacs,superscriptaddress]{revtex4-1}
\usepackage{amsmath}
\usepackage{epsfig}
\usepackage{color}
\usepackage{endnotes}
\usepackage{natbib}
\usepackage[colorlinks,breaklinks]{hyperref}
\usepackage{nicefrac}
\usepackage{bm}
\usepackage{dcolumn}
\usepackage{graphics}
\usepackage{graphicx} 
\usepackage{tikz} 
\usepackage{amsmath}
\usepackage{amssymb}
\usepackage{color}
\usepackage{changes}
\usepackage{multirow}
\usepackage{lineno}

\newcommand{\bs}[1]{\ensuremath{\boldsymbol{#1}}}

\newcommand{\be}{\begin{equation}}
\newcommand{\ee}{\end{equation}}
\newcommand{\bea}{\begin{align}}
\newcommand{\eea}{\end{align}}
\newcommand{\beqa}{\begin{eqnarray}}
\newcommand{\eeqa}{\end{eqnarray}}

\newcommand*{\MIT }{Massachusetts Institute of Technology, Cambridge, Massachusetts 02139, USA}
\newcommand*{\ODU}{Old Dominion University, Norfolk, Virginia 23529, USA}

\newcommand*{\TAU }{School of Physics and Astronomy, Tel Aviv University, Tel Aviv 69978, Israel}

\newcommand*{\GW}{George Washington University, Washington, D.C., 20052, USA}
\newcommand*{\HUJI}{The Racah Institute of Physics, The Hebrew University, Jerusalem 9190401, Israel}

\begin{document}



\title{Extracing the number of short-range corerlated nucleon pairs from inclusive
electron scattering data}

\author{R. Weiss}
\thanks{Equal Contribution}
\affiliation{\HUJI}
\author{A.W. Denniston}
\thanks{Equal Contribution}
\affiliation{\MIT}
\author{J.R. Pybus}
\affiliation{\MIT}

\author{O. Hen}
\affiliation{\MIT}
\author{E. Piasetzky}
\affiliation{\TAU}
\author{A. Schmidt}
\affiliation{\GW}
\author{L.B. Weinstein}
\affiliation{\ODU}

\author{N. Barnea}
\email[Contact Author \ ]{(nir@phys.huji.ac.il)}
\affiliation{\HUJI}

\begin{abstract} 
  The extraction of the relative abundances of short-range correlated
  (SRC) nucleon pairs from inclusive electron scattering is studied
  using the generalized contact formalism (GCF) with several nuclear
  interaction models.  GCF calculations can reproduce the observed
  scaling of the cross-section ratios for nuclei relative to deuterium
  at high-$x_B$ and large-$Q^2$, $a_2=(\sigma_A/A)/(\sigma_d/2)$.  In
  the non-relativistic instant-form formulation, the calculation is
  very sensitive to the model parameters and only reproduces the data
  using parameters that are inconsistent with ab-initio many-body
  calculations.  Using a light-cone GCF formulation significantly
  decreases this sensitivity and improves the agreement with ab-initio
  calculations.  The ratio of similar mass isotopes, such as $^{40}$Ca
  and $^{48}$Ca, should be sensitive to the nuclear asymmetry
  dependence of SRCs, but is found to also be sensitive to low-energy
  nuclear structure.  Thus the empirical association of SRC pair
  abundances with the measured $a_2$ values is only accurate to about
  $20\%$.  Improving this will require cross-section calculations that
  reproduce the data while properly accounting for both nuclear structure
  and relativistic effects.
\end{abstract}

\maketitle

To a good approximation, neutrons and protons
with momentum below the Fermi sea can be considered as 
independently moving in well-defined quantum orbits of the average,
mean-field, nuclear interaction. Above the Fermi sea, 
short-range correlated (SRC) pairs
dominate \cite{Hen:2016kwk,Atti:2015eda,
piasetzky06,subedi08,alvioli08,hen14,korover14,Weiss:2016obx,
Duer:2018sxh,Korover:2020lqf}.
Therefore, quantifying the number of correlated pairs
is important for obtaining a complete picture of the atomic nucleus.

A description of correlations in complex nuclear systems 
can be done in the spirit of the successful atomic theory,
in which various properties of a unitary gas
are connected to a single parameter, the
{\it contact} \cite{Tan08a,Tan08b,Tan08c,Braaten12}.
In essence, the contact counts the number of SRC pairs in the system.
The importance of this quantity to nuclear systems
was demonstrated by the success of the 
generalized contact formalism (GCF),
which takes into account the complicated nature of the nuclear 
force~\cite{Weiss14,Weiss:2015mba,Weiss:2016obx,
Cruz-Torres:2017sjy,Cruz-Torres:2019fum, Weiss:2018tbu,schmidt20}.
SRC pair abundances are also used in modeling the
effective impact of SRCs on the nuclear symmetry energy and
neutron-star properties~\cite{hen15,Cai:2015xga,Li:2018lpy,Souza:2020gjs}, and in
studies of the modification of quark distributions in
nuclei~\cite{Hen:2016kwk,weinstein11,Hen12,Hen:2013oha,Chen:2016bde,Ciofi07},
the flavor dependence of the EMC
effect~\cite{Schmookler:2019nvf,Arrington:2019wky,Hen:2019jzn}, and
low-energy QCD symmetry breaking
mechanisms~\cite{hen11,Segarra:2019gbp}.

Inclusive electron scattering $(e,e')$ measurements are commonly used to estimate 
SRC pair abundances in nuclei. In kinematics sensitive to SRCs,
the cross-section ratio $\sigma_A/A\;/\;\sigma_d/2$, between
nucleus $A$ and the deuterium
``scales'', reaching a constant
value independent of the momentum and energy 
transfer~\cite{frankfurt93,egiyan02,egiyan06,fomin12,Schmookler:2019nvf,Nguyen:2020mgo}.
The value of this constant, $a_2(A/d)$ or simply $a_2$, is traditionally
interpreted as the number of
neutron-proton ($np$) deuteron-like SRC pairs in nucleus $A$ relative to
deuterium~\cite{frankfurt93,egiyan02,egiyan06,fomin12,Schmookler:2019nvf,Hen:2016kwk,Nguyen:2020mgo}.

This scaling is seen at kinematic of $Q^2\gtrsim 1.4$ GeV$^2$ and $1.5\le x_B \le 1.9$,
where $x_B=Q^2/2m\omega$, $Q^2=\bs{q}^2 -\omega^2$, $\bs{q}$ and
$\omega$ are the three-momentum and energy transfer respectively, and
$m$ is the nucleon mass.
The value of $x_B \ge 1.5$ determines that the 
minimum allowed initial momenta $k_{min}$ of the struck nucleon
is very close to the typical nuclear Fermi momentum for medium to heavy nuclei, $k_{F}
\approx 250$ MeV/c \cite{egiyan02}.
Nucleons with higher momenta are predominantly
part of deuteron-like SRC
pairs~\cite{piasetzky06,subedi08,alvioli08,korover14,hen14,Weiss:2016obx,Duer:2018sxh,Korover:2020lqf}.
The scaling then naturally arises in a simplistic SRC picture where
the struck nucleon belongs to a stationary deuteron-like pair.  
In this picture the recoil momentum is carried by a 
single nucleon and the $A-2$ residual nucleus does not recoil.  
Therefore, $k_{min}$ of the struck nucleon
and its ground-state momentum distribution
 are similar in deuterium and heavier nuclei, 
resulting in cross-section ratio scaling
that should be proportional to the number
of SRC pairs~\cite{frankfurt93,egiyan02}.

\begin{figure}[t]
\begin{center}
\includegraphics[height=5.5 cm]{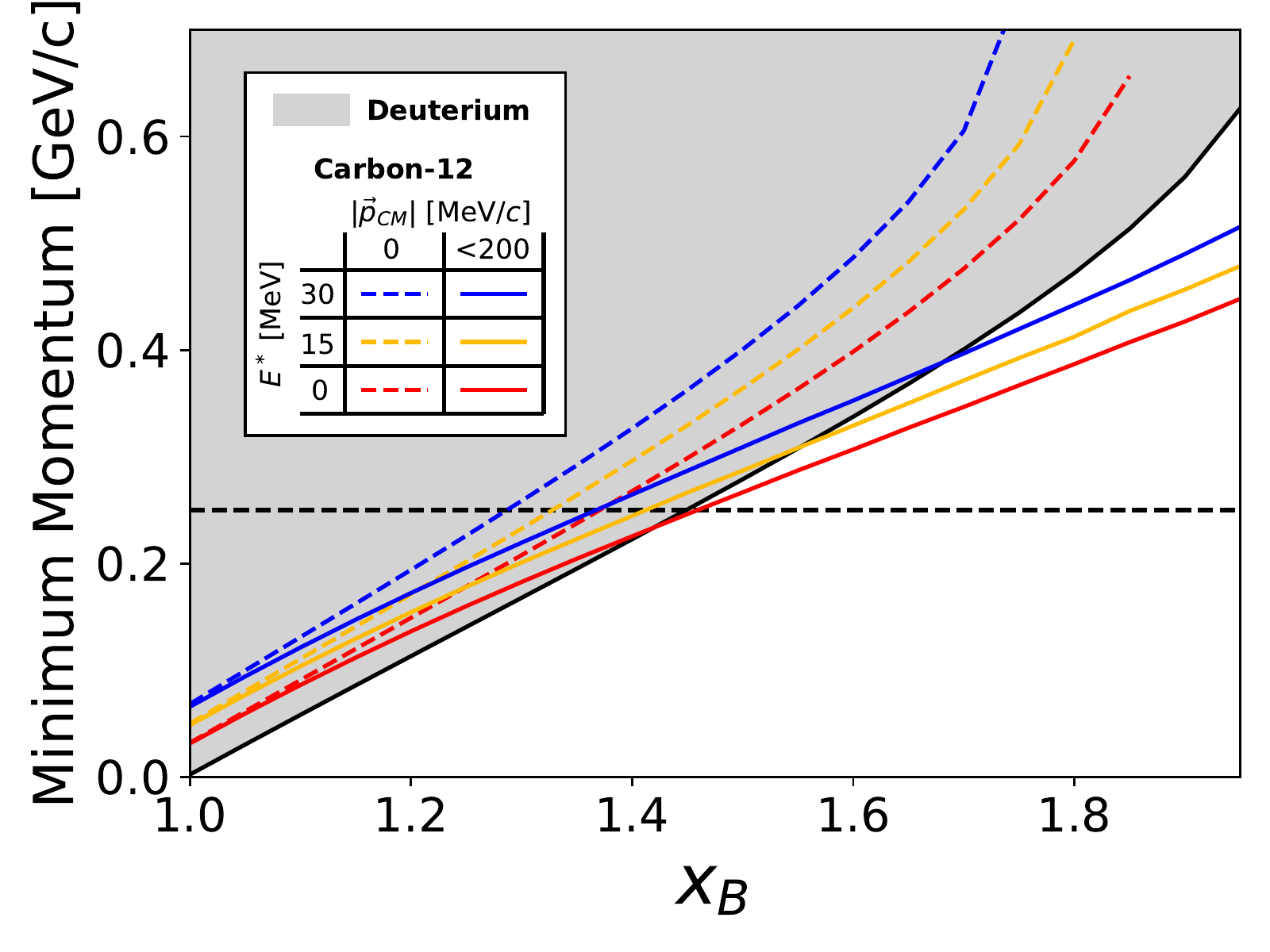}
	\caption{
	The minimum possible momentum of the nucleon absorbing the
	virtual photon, $k_{min}$, in inclusive scattering as a function of $x_B$,
	for $Q^2=2$ GeV$^2$.  The black line shows $k_{min}$ for the
        deuteron, while the colored lines show $k_{min}$ for SRC pairs in $^{12}$C, for
        different $A-2$ excitation energies, $E^*_{A-2}$, and for different
        pair center-of-mass momenta, denoted by $|\vec{p}_{CM}|$. 
        The gray region shows the initial momentum
        range, $k\ge k_{min}$, for $d(e,e')$.  The horizontal dashed line
	corresponds to the Fermi momentum for heavy nuclei, $k_F \approx 0.25$ GeV/c.
	}
\label{Fig:p_min}
\end{center}
\end{figure}

However, this intuitive interpretation of $a_2$ in terms of SRC
abundances neglects important effects: (1) the presence of
non-deuteron-like SRCs (proton-proton $(pp)$, neutron-neutron $(nn)$,
and $pn$ pairs with $s\neq1$), (2) pair center-of-mass (CM)
motion~\cite{Cohen:2018gzh}, and (3) possible excitation of the
residual $A-2$ system.  CM motion and $A-2$ excitation can
dramatically affect $k_{min}$ (see Fig.~\ref{Fig:p_min}) which can
significantly affect the simplistic interpretation of $a_2$.

In addition, final-state interaction (FSI) can contribute to the
measured $(e,e')$ cross-sections and disrupt this simplistic
interpretation of $a_2$.  While such contributions grow with $x_B$ and
can reach up to $50\%$, it was argued by several
calculations~\cite{Frankfurt81,Frankfurt88,frankfurt93,CiofidegliAtti:1994ys,CiofidegliAtti:1995qe,Benhar95,frankfurt08b}
(but not all~\cite{Benhar95}) that they are confined to within SRC
pairs and cancel to first approximation in the $A/d$ ratio. The main
input for the FSI calculations are measured $NN$ scattering
cross-sections and these calculations are done in a high-resolution
reaction model using one-body reaction operators, similar to the
reaction scheme employ by our GCF calculations.

As more and better $a_2$ data are becoming available
~\cite{Fomin:2017ydn}, and as studies utilizing $a_2$ values as
SRC abundances demand higher precision~\cite{Nguyen:2020mgo},
it is timely to examine the quantitative impact of realistic SRC
modeling on the classical interpretation of $a_2$.
Such modeling is also important for establishing a direct connection
between inclusive electron scattering
and ab-initio many-body structure
calculations~\cite{schiavilla07,alvioli08,Feldmeier:2011qy,neff15,carlson12,wiringa14,Lynn:2019vwp,Cruz-Torres:2019fum}.

Here we study the interpretation of $a_2$ scaling using the
GCF to calculate high-$x_B$ high-$Q^2$
inclusive scattering cross-sections.  By comparing measured and
GCF-calculated cross-sections using different model parameters we
provide a new, quantitative, understanding of the model dependence of
SRC pair abundance extraction.

The GCF is a realistic effective model of SRCs, used to connect
experimental data and ab-initio nuclear structure
calculations~\cite{Weiss:2015mba,Weiss:2016obx,Cruz-Torres:2019fum}.
Building on the scale-separation of nucleons in SRC pairs from the
surrounding nuclear environment, it models nucleons in SRC pairs using
universal (i.e., nucleus independent) two-particle functions, and
system- and state-dependent \emph{contact} terms that describe the
abundance of SRC pairs.  This scale-separated approach 
successfully reproduced ab-initio calculated nucleon distributions at
short-distance and high-momentum, enabling a meaningful extraction of
nuclear contact
terms~\cite{Weiss:2015mba,Weiss:2016obx,Cruz-Torres:2019fum}.  More
recently, it was extended to model nuclear spectral and correlation
functions~\cite{Cruz-Torres:2017sjy,Weiss:2018tbu}, enabling a
successful reproduction of a wide range of $(e,e'N)$ and $(e,e'NN)$
measurements~\cite{Weiss14,Weiss:2016obx,Weiss:2018tbu,schmidt20,Duer:2018sxh,Pybus:2020itv}.  
The GCF thus provides an established and robust formalism to describe
experimental data using effective parameters obtained from many-body
calculations.


To quantify the impact of these effects we perform GCF calculations of
inclusive cross-section ratios using various parameters and compare
them to each other and to experimental data.  We used both
non-relativistic instant-form (IF) and light-cone (LC) GCF
formulations, to see the effect of relativistic corrections for these
high-momentum nucleons.  We integrated the previously derived the GCF
$(e,e'N)$ and $(e,e'NN)$ cross-sections over the knocked-out nucleons,
to obtain the inclusive $(e,e')$ cross-section.

Within the plane-wave impulse approximation (PWIA), the IF GCF
$(e,e'NN)$ cross-section for the breakup of an SRC pair is given
by~\cite{Pybus:2020itv}
\begin{equation}
\begin{split}
&\frac{d^8\sigma_A}{dE_e d\Omega_e d^3\vec{p}_\text{CM} d\Omega_\text{rel}}  \\
&\quad=\kappa_{IF}  
\sum_{N_1N_2,\beta} s\sigma_{eN_1} C^{A,\beta}_{N_1N_2}|\tilde{\varphi}^\beta_{N_1N_2}(\vec{p}_\text{rel})|^2
n^{A,\beta}_{N_1N_2}(\vec{p}_\text{CM})\\
&\quad\equiv\sum_{N_1N_2,\beta} C^{A,\beta}_{N_1N_2} \times \sigma_{N_1N_2,IF}^{\beta},
\label{eq:finalNR}
\end{split}
\end{equation}
where $E_e$ and $\Omega_e$ are the energy and solid angle of the
scattered electron, and $\vec{p}_\text{CM}$ and $\vec{p}_\text{rel}$
are the CM and relative momenta of the initial-state SRC pair,
respectively.  $\sigma_{eN_1}$ is the off-shell electron-nucleon cross
section, $s$ is a symmetry factor ($s=1$ for $np$ and $pn$ and
$s=2$ for $nn$ and $pp$), and $\kappa_{IF} \equiv \frac{1}{32
  \pi^4} \frac{p_\text{rel}^3 E_1' E_2}{\left|
    \left(E_2\vec{p}_1'+E_1'\vec{p}_2\right)\cdot\vec{p}_\text{rel}\right|}$
is a phase-space factor, where $(\vec{p}_1',E_1')$ and $(\vec{p}_2,E_2)$
are the knocked-out and spectator nucleon four-momenta, respectively.
$|p_{rel}|$ is fixed by energy-momentum conservation.

$C_{N_1N_2}^{A,\beta}$ are nucleus-dependent nuclear contacts,
measuring the probability to find an $N_1N_2$ SRC pair ($pp$, $nn$, $np$ or $pn$)
 in nucleus $A$ with quantum numbers $\beta$. 
$\beta=1$ denotes spin-one deuteron-like pairs, and $\beta=0$ is for the spin-zero $s$-wave pairs.
$n_{N_1N_2}^{A,\beta}(\vec{p}_\text{CM})$ is the SRC pairs CM momentum distribution, 
approximated by a three-dimensional Gaussian with an $A$-dependent width $\sigma_{CM}$~\cite{Cohen:2018gzh,CiofidegliAtti:1995qe,Colle:2013nna}.
${\tilde{\varphi}_{N_1N_2}^{\beta}}$ are the universal two-body functions
of the relative momentum distribution of nucleons in SRC pairs,
obtained by solving the zero-energy two-body Schr\"odinger equation with
a given $NN$ interaction model (e.g., AV18, N2LO etc.).

We stress that the contact values are fixed by
comparison with ab-initio calculations~\cite{Cruz-Torres:2019fum} and
$\sigma_{CM}$ was measured in Ref.~\cite{Cohen:2018gzh}. 
The unmeasured average excitation energy of the residual system
$E^*_{A-2}$ is limited by the
typical excitation energy of the system ($0\le E^*_{A-2}\le 30$ MeV).
The unceratinties of these parameters are used to evaluate the
uncertainties of the GCF calculations.

Light-cone four-momentum vectors are expressed in terms of
longitudinal (along the $\bs{q}$ direction) plus- and minus-momentum
$p^\pm \equiv p^0 \pm p^3$ and transverse momentum $\vec{p}^\perp
\equiv (p^1,p^2)$.  The light-cone momentum fraction is $\alpha \equiv
p^-/\bar{m}$, where $\bar{m} = m_A/A$.
The advantages of studying inclusive reactions using LC are discussed in~\cite{frankfurt93}.

The PWIA LC GCF $(e,e'NN)$ cross section is given by~\cite{Pybus:2020itv}
\begin{equation}
\begin{split}
&\frac{d^8\sigma_A}{dE_ed\Omega_ed^3\vec{p}_\text{CM}d\Omega_{rel}}  = \sum_\beta C^{A,\beta}_{N_1N_2} \times \sigma_{N_1N_2,LC}^{\beta},
\label{eq:finalLC}
\end{split}
\end{equation}
where
\begin{equation}
\begin{split}
\sigma_{N_1N_2,LC}^{\beta} = s\kappa_{LC}  
 \sigma_{eN_1} \psi^\beta_{N_1N_2}(\alpha_\text{rel},\vec{p}^\perp_\text{rel})\rho^{A,\beta}_{N_1N_2}(\alpha_\text{CM},\vec{p}^\perp_\text{CM}).
\label{eq:finalLC_extended}
\end{split}
\end{equation}
Here $\alpha_\text{CM}$, $\vec{p}^\perp_\text{CM}$, $\alpha_\text{rel}$, and $\vec{p}^\perp_\text{rel}$
are the LC longitudinal, LC transverse, CM, and relative momenta of the SRC pair, respectively.
$\kappa_{LC} = \kappa_{IF}  \frac{8 \pi^3 \alpha_{A-2}}{\alpha_1 \alpha_\text{CM} E_{A-2}}$
is a phase-space factor.
$\rho^{A,\beta}_{N_1N_2}(\alpha_\text{CM},\vec{p}_\text{CM})$ is a
three-dimensional gaussian of width $\sigma_\text{CM}$ and 
$
\psi^\beta_{N_1N_2}(\alpha_\text{rel},\vec{p}^\perp_\text{rel})= \frac{\sqrt{m_N^2+k^2}}{2-\alpha_\text{rel}}
\frac{|\tilde{\varphi}^\beta_{N_1N_2}(k)|^2}{(2\pi)^3}
$
is the LC equivalent of the IF universal function~\cite{piasetzky06}
where $k=\frac{m^2+k_\perp^2}{\alpha_{rel}(2-\alpha_{rel})}-m^2$.

\begin{figure}[t]
	\centering  
	\includegraphics[width=\columnwidth,height=6.5cm]{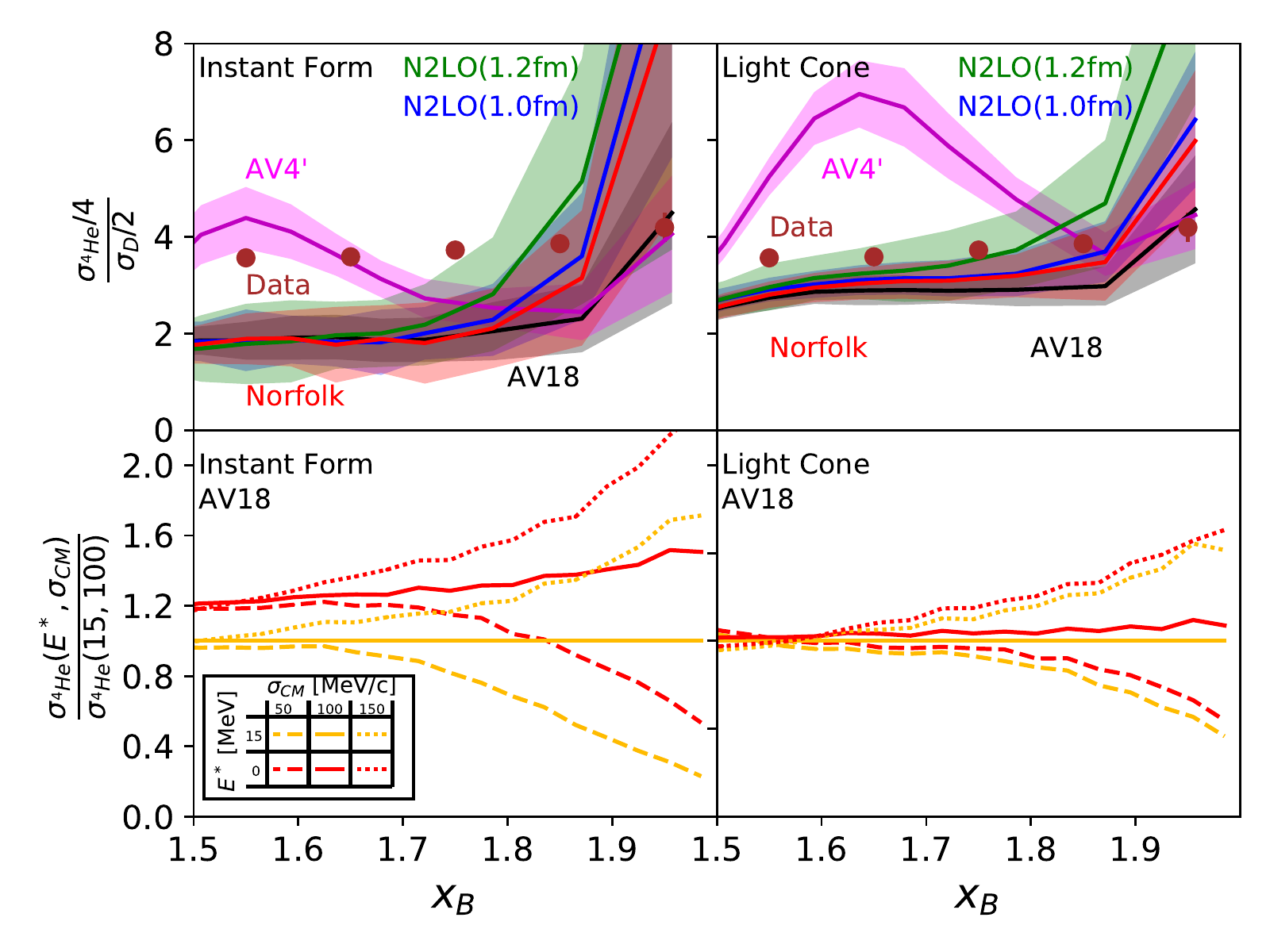}
	\caption{Top: Measured per-nucleon $(e,e')$ cross-section ratios $\sigma_{^4He}/4\;/\;\sigma_d/2$ as a function of $x_B$. 
          The data~\cite{fomin12} are compared with GCF calculations
          using both instant form (left) and light cone (right) GCF formulations with different $NN$ interaction models
          and using $\sigma_{CM} = 100\pm20$ MeV/c~\cite{korover14,Cohen:2018gzh}, 
          $E^*_{A-2} = 0 - 30$ MeV, and contact parameters from Ref.~\cite{Cruz-Torres:2019fum}.
          The width of the bands show their 68\% confidence
          interval due to the uncertainties in the model parameters. 
          Bottom: Ratio of the GCF calculated $^4$He cross section with different excitation
          energies ($E^*_{A-2}$) and CM momentum distribution widths ($\sigma_{CM}$) to the cross section
          calculated for $E^*_{A-2}=15$ MeV and $\sigma_{CM}=100$ MeV/c.
          Calculations were done using both instant form (left) and light cone (right) GCF formulations
          with AV18~\cite{wiringa95}  $NN$ interaction model.}
	\label{fig:4He_GCF}
\end{figure}

By integrating Eqs. \eqref{eq:finalNR} or \eqref{eq:finalLC}
we obtain the IF or LC GCF inclusive cross section:
\begin{equation}
\label{inclusive_cross_section}
\frac{d^3\sigma_A}{dE_{k'} d\Omega_{k'}}
=\sum_{N_1N_2, \beta}  C^{A,\beta}_{N_1N_2} \int \sigma_{N_1N_2}^{\beta} \; d^3\vec{p}_\text{CM}d\Omega_\text{rel}  ,
\end{equation}
where the sum spans $s=1$ $np$-SRC and $s=0$ $np$-, $pp$- and $nn$-SRC pairs 
and includes the electron coupling to either nucleon of the pair.  
The integration is limited by energy-momentum conservation and
depends on $\sigma_{CM}$ and $E^*_{A-2}$.

For the simple case of interacting with standing (i.e. no pair c.m.~motion) on-shell (i.e. no $E^*_{A-2}$ effects) SRC pairs,
the cross-section ratio for nucleus $A$ relative to deuterium is given by:
 \begin{equation}
 \label{inclusive_cross_section_Ratio}
 \begin{split}
 \frac{\sigma_A}{\sigma_d} =& \frac{C^{A,s=1}_{pn} }{C^{d,s=1}_{pn}} \times 
  \left[ 
 1 + 
 \frac{C^{A,s=0}_{pn}}{C^{A,s=1}_{pn}} \frac{{\Psi}^{s=0}_{pn}}{{\Psi}^{s=1}_{pn}}+
 2\frac{C^{A,s=0}_{pp}}{C^{A,s=1}_{pn}} \frac{{\Psi}^{s=0}_{pp}}{{\Psi}^{s=1}_{pn}} \right] ,
 \end{split}
 \end{equation}
 where the factor of $2$ before the $pp$ term accounts for $nn$ pairs assuming isospin symmetry
 and ${\Psi}^\beta_{N_1N_2}$ represent the phase-space integral over the 
 universal functions $\tilde{\varphi}^\beta_{N_1N_2}$ (instant form) or 
 $\tilde{\psi}^\beta_{N_1N_2}$ (light-cone).
 As $\frac{C^{A,s=0}_{N_1N_2}}{C^{A,s=1}_{pn}} \ll 1$ for any $NN$ interactions with a tensor force 
 (for all $N_1N_2$ pairs), the cross-section ratio in this simplistic case will approximately equal 
 $\frac{C^{A,s=1}_{pn} }{C^{d,s=1}_{pn}}$. The latter was previously shown~\cite{Cruz-Torres:2019fum} to be insensitive
 to the $NN$ interaction model. It is thus expected for the $A/d$ cross-section ratio
 to be dominated by mean-field properties of the nucleus and thus be largely insensitive to the
 $NN$ interaction model~\cite{Chen:2016bde,Lynn:2019vwp,Ryckebusch:2019oya,Cruz-Torres:2019fum}.

Fig.~\ref{fig:4He_GCF} (top panels) shows the measured~\cite{fomin12} and
GCF-calculated $\sigma_{^4He}/4\;/\;\sigma_D/2$ cross-section ratio,
using nuclear contacts and c.m. width from refs.~\cite{korover14,Cohen:2018gzh,Cruz-Torres:2019fum}, 
$E^*_{A-2} = 0 - 30$ MeV, and universal functions calculated with 
several $NN$ interaction models, including the phenomenological AV18~\cite{wiringa95} and AV4'~\cite{Wiringa:2002},
and the chiral NV2+3-Ia* (Norfolk)~\cite{Piarulli:2016vel, Piarulli:2017dwd, Baroni:2018fdn} and N2LO~\cite{Gezerlis:2014,Lynn:2016,Lonardoni:2018prc} interaction with $1.0$ and $1.2$ fm cutoffs.  
Both IF and LC ratios show scaling plateaus (i.e. are constant for $1.4\le x_B\le 1.9$), 
but the IF ratio is almost a factor of two too low.
Calculations for additional nuclei are shown in the supplementary materials.

The calculations are largely insensitive to the $NN$ interaction
model, except for the special case of AV4' which does not include a
tensor force and is therefore not dominated by deuteron-like pairs.
This sensitivity of the GCF calculation to the Tensor force stands in
contrast with the EFT analysis of Ref.~\cite{Lynn:2019vwp} where the
calculation does not directly employ high-resolution one-body reaction
operators and the nature of the two-body interaction completely
cancels in the cross-section ratio.

The marginal performance of the IF calculations is very surprising as
they reproduce $(e,e'N)$ and $(e,e'NN)$ data at similar kinematics
remarkably well~\cite{schmidt20,Pybus:2020itv}.  The LC ratios are
better, but are still $\approx25\%$ lower than the data.
This might point to an issue with the contact extraction from ab-initio calculations,
becasue the results of Ref.~\cite{schmidt20,Pybus:2020itv} are not sensitive to
the $A/d$ contact ratio. 
In the LC case, a 10--20\% relativistic correction to the contact extraction could
explain the data.

\begin{figure}[t]
	\centering  
	\includegraphics[width=\linewidth]{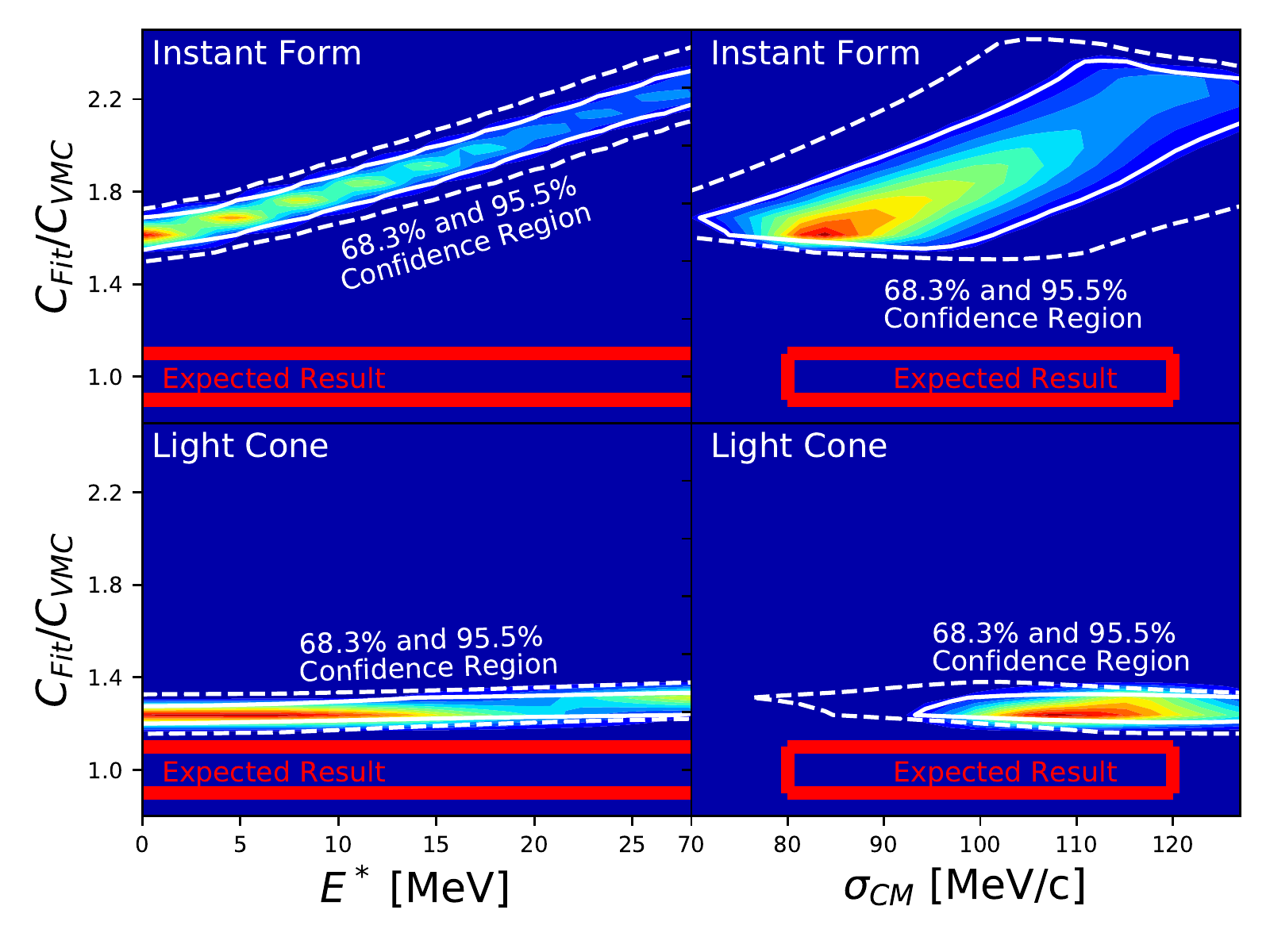}
	\caption{ GCF parameter confidence intervals for fitting
          $^4$He$(e,e') / d(e,e')$ data of ref.~\cite{fomin12} using
          instant form (top) and light cone (bottom) GCF formulations
          with the AV18 $NN$ interaction~\cite{wiringa95}. The color scale
          represents the likelihood of the fit parameters given the data,
          with the white solid (dashed)
          contours indicating the 68.3\% (95.5\%) confidence regions. 
          Red lines show the expected parameter values from previous
          measurements and/or ab-initio 
          calculations~\cite{Cruz-Torres:2019fum}. The contact value
          $C_{np}^{s=1}$ is shown as a ratio to its value extracted
          from many-body Variational Monte-Carlo (VMC) calculations.
          See text for details.  }
	\label{fig:4He_Fit}
\end{figure}

\begin{figure*}[t]
	\centering  
	\includegraphics[width=0.75\linewidth]{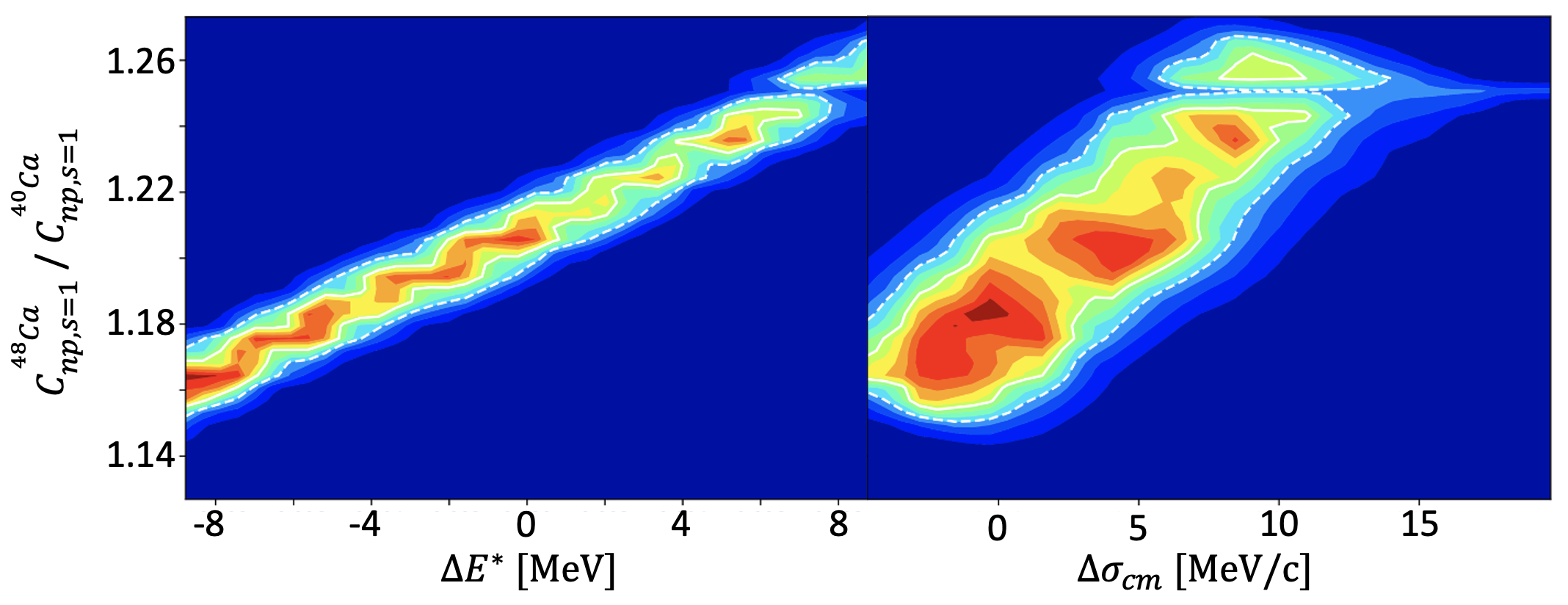}
	\caption{Likelihood map for the correlation between the
          extracted ratio of spin-1 $pn$ contacts in $^{48}$Ca over
          $^{40}$Ca, $C_{pn,s=1}^{48Ca} / C_{pn,s=1}^{40Ca}$, and
          $\Delta E^* = E^*_{46K} - E^*_{38K}$ (left) and $\Delta
          \sigma_{CM} = \sigma_{CM}^{48Ca} - \sigma_{CM}^{40Ca}$
          (right).  The parameters likelihoods are determined by
          fitting the $^{40}$Ca$(e,e')$ / $^{48}$Ca$(e,e')$
          cross-section ratio data of Ref.~\cite{Nguyen:2020mgo} with
          GCF calculations in the $x_B$ range of $1.5 \leq x_B \leq
          1.9$.  The calculation used the GCF light-cone formulations
          with the AV18 $NN$ interaction~\cite{wiringa95}.  The color
          scale represents the likelihood of the fit parameters given
          the data, with the white solid (dashed) contours indicating
          the 68.3\% (95.5\%) confidence regions.  See text for
          details.}
	\label{fig:Ca4840_Fit}
\end{figure*}

To better understand this discrepancy we examined the impact
of varying $\sigma_{CM}$ by $\pm50$ MeV/c and $E^*_{A-2}$  from 0 to
30 MeV using the AV18 interaction (see Fig.~\ref{fig:4He_GCF} (bottom)).
The IF calculation is very sensitive to both parameters.
A 15 MeV change in $E^*_{A-2}$ changes the cross-section by $\sim 20\%$.
A 50 MeV/c change in $\sigma_{CM}$ changes the cross section dramatically starting at $x_B=1.7$.  
Ref.~\cite{vanhalst12} also predicted large effects (up to 70\%) due to pair CM motion,
which is very different than the $19\pm6\%$ $x_B$-independent correction used 
by Ref.~\cite{fomin12}, motivated by a simplistic one-dimensional gaussian smearing
of the deuteron momentum distribution~\cite{Arrington12}.  

This sensitivity indicates that different effects, such as $A$-dependent
FSIs~\cite{Benhar95}, or contributions from $3N$-SRCs that are missing
in the current GCF calculations and are estimated to be a $~\sim 10\%$
correction to the leading $2N$-SRC
contribution~\cite{egiyan06,Weiss:2016obx,Sargsian:2019joj,Korover:2020lqf},
might explain the disagreement seen in Fig. \ref{fig:4He_GCF}.
The study of such corrections is ongoing and
extends beyond the scope of the present work.
It also raises concerns about the ability to study the mass
and asymmetry dependence of SRC pair abundances using pairs abundances
extracted from $(e,e')$ measurements of light nuclei where
$\sigma_{CM}$ and $E^*_{A-2}$ vary significantly.

Lastly we studied what parameter values are needed to
describe the data.  
We varied $\sigma_{CM}$, $E^*_{A-2}$, and
the spin-1 contact ratio $C_{np}^{A,s=1}/C_{np}^{d,s=1}$, to fit the
$^4$He/d~\cite{fomin12} and $^{12}$C/d data~\cite{Schmookler:2019nvf}.  
We kept the $C_{np}^{s=1}/C_{NN}^{s=0}$ ratio fixed.
The IF and LC results  both described the data well.
See online supplementary materials for details.

The resulting $^4$He AV18 parameters and their
correlations are shown in Fig.~\ref{fig:4He_Fit}.  Results for the
other $NN$ interaction models and different nuclei are shown in the
online supplementary materials Table I.
The fitted contacts have much larger uncertainties (up to $30\%$
for IF and just under $10\%$ for LC) 
than the typical $2\%$ experimental uncertainties in $a_2$.  For the
LC case this comes primarily from $\sigma_{CM}$, but  IF is also sensitive
to 
$E^*_{A-2}$.  

The fitted IF contact ratios for deuteron-like $np$ pairs are higher
than the VMC calculation results by $50 - 150\%$ for both $NN$
interactions and both nuclei, as expected from the results of
Fig.~\ref{fig:4He_GCF}.  The fitted LC contacts are only $20 - 30\%$ higher
than the VMC calculations for both $NN$ interactions, which is
not much more than the $\sim10\%$ uncertainties on both the calculated
and fitted contacts.  For $^{12}$C the same holds true for AV18 but a
larger $80\%$ disagreement is observed for N2LO.

Comparing with $a_2$, that are traditionally interpreted as a measure
of deuteron-like $np$ pairs, the fitted values are within $10 - 15$\%
of the data for both $^4$He and $^{12}$C, except for IF N2LO, which is
within
$\sim 30\%$.  However, this is an accidental result of the
cancellation between the effects of $\sigma_{CM}$ and the contribution
of non-deuteron-like pairs, which increase the ratio, and the effect
of $E^*_{A-2}$, which decreases the ratio. This
cancellation should be quite different in light and asymmetric nuclei
where $\sigma_{CM}$, $E^*_{A-2}$ and the $np/pp$-pair ratio can
change rapidly with $A$.

To examine the effect of the nuclear asymmetry, we analyzed recent
measurements of
$a_2(^{48}\mathrm{Ca}/^{40}\mathrm{Ca})$~\cite{Nguyen:2020mgo}.  The
calculation used $^{40}$Ca contacts from
Ref.~\cite{Cruz-Torres:2019fum} and assumed the same spin-0 contact
for $^{48}$Ca.  We varied the spin-1 $^{48}$Ca contact and the values of
$E^*_{A-2}$ and $\sigma_{CM}$ for each nucleus.

The calculation was relatively insensitive to 
$E^*_{A-2}$ and $\sigma_{CM}$. However, it could not place a stringent constraint on the
important $^{48}$Ca/$^{40}$Ca spin-1 contact ratio, because that
is extremely sensitive to the
parameter differences between $^{48}$Ca and
$^{40}$Ca, $\Delta \sigma_{CM} = \sigma_{CM}^{48Ca} -
\sigma_{CM}^{40Ca}$ and $\Delta E^* = E^*_{46K} - E^*_{38K}$
(see Figure~\ref{fig:Ca4840_Fit}).   A 10 MeV change in either
parameter difference induces a large change in the extracted contact
ratio.  This few-MeV nuclear structure difference could plausibly be
caused by the
neutron-skin of $^{48}$Ca and the very different energy
levels of $^{38}$K and $^{46}$K.

This again emphasizes the large model dependence of interpretations of
the measured
nuclear asymmetry dependence of $a_2$, even in similar
mass nuclei.  This has direct implications for studies that use
the asymmetry dependence of $a_2$, e.g., for understanding the flavor
dependence of the EMC effect~\cite{Arrington:2019wky} and the
properties of nucleons in dense neutron-rich
matter~\cite{sargsian14,McGauley:2011qc,hen15,Li:2018lpy,Souza:2020gjs}.

For completeness we note that the inclusive cross-sections can also be analyzed in a
complementary low-resolution picture with many-body operators and no
SRCs~\cite{More:2017syr}.  This has not been implemented in the GCF
and goes beyond the scope of the current work.  In addition,
calculations in Effective Field Theory (EFT) approximate $a_2$ using
the ratio of two-nucleon densities at short distance for nucleus $A$
and the deuteron~\cite{Chen:2016bde,Lynn:2019vwp}.  This approach
reproduces $a_2$ values, but cannot model the $x_B$ or $Q^2$
dependences of the ratio or provide insight into specific pair
characteristics such as $\sigma_{CM}$ and the relation between $a_2$
values and low-energy nuclear structure (i.e. impact of $E^*_{A-2}$).


To conclude, $a_2$ measurements are widely used to extract SRC abundances,
with wide ranging implications.
Our calculations suggest that the traditional interpretation of $a_2$ as an empirical
measure of the abundance of deuteron-like $np$-SRC pairs in
nucleus $A$ relative to the deuteron is accurate to about $20\%$.  
This has significant implications for planned precision measurements~\cite{Fomin:2017ydn} 
of the nuclear mass and asymmetry dependence of $a_2$, especially for light nuclei.  
While the cross section ratio $a_2$ can be measured precisely, 
supplemental $(e,e'N)$ and $(e,e'NN)$ measurements and detailed 
cross section calculations are needed for its accurate interpretation.

\begin{acknowledgments}
We thank M. Strikman, M. Sargsian, W. Cosyn, J. Ryckebusch and C. Weiss for insightful discussions.
This work was supported by the U.S. Department of Energy, Office of Science, Office of Nuclear Physics under Award Numbers DE-FG02-94ER40818, de-sc0020240, DE-FG02-96ER-40960, DE-FG02-93ER40771, and DE- AC05-06OR23177 under which Jefferson Science Associates operates the Thomas Jefferson National Accelerator Facility, the Israeli Science Foundation (Israel) under Grants Nos. 136/12 and 1334/16, the Pazy foundation, and the Clore Foundation.
\end{acknowledgments}

\bibliography{../../../../references.bib}

\begin{thebibliography}{70}%
\makeatletter
\providecommand \@ifxundefined [1]{%
 \@ifx{#1\undefined}
}%
\providecommand \@ifnum [1]{%
 \ifnum #1\expandafter \@firstoftwo
 \else \expandafter \@secondoftwo
 \fi
}%
\providecommand \@ifx [1]{%
 \ifx #1\expandafter \@firstoftwo
 \else \expandafter \@secondoftwo
 \fi
}%
\providecommand \natexlab [1]{#1}%
\providecommand \enquote  [1]{``#1''}%
\providecommand \bibnamefont  [1]{#1}%
\providecommand \bibfnamefont [1]{#1}%
\providecommand \citenamefont [1]{#1}%
\providecommand \href@noop [0]{\@secondoftwo}%
\providecommand \href [0]{\begingroup \@sanitize@url \@href}%
\providecommand \@href[1]{\@@startlink{#1}\@@href}%
\providecommand \@@href[1]{\endgroup#1\@@endlink}%
\providecommand \@sanitize@url [0]{\catcode `\\12\catcode `\$12\catcode
  `\&12\catcode `\#12\catcode `\^12\catcode `\_12\catcode `\%12\relax}%
\providecommand \@@startlink[1]{}%
\providecommand \@@endlink[0]{}%
\providecommand \url  [0]{\begingroup\@sanitize@url \@url }%
\providecommand \@url [1]{\endgroup\@href {#1}{\urlprefix }}%
\providecommand \urlprefix  [0]{URL }%
\providecommand \Eprint [0]{\href }%
\providecommand \doibase [0]{http://dx.doi.org/}%
\providecommand \selectlanguage [0]{\@gobble}%
\providecommand \bibinfo  [0]{\@secondoftwo}%
\providecommand \bibfield  [0]{\@secondoftwo}%
\providecommand \translation [1]{[#1]}%
\providecommand \BibitemOpen [0]{}%
\providecommand \bibitemStop [0]{}%
\providecommand \bibitemNoStop [0]{.\EOS\space}%
\providecommand \EOS [0]{\spacefactor3000\relax}%
\providecommand \BibitemShut  [1]{\csname bibitem#1\endcsname}%
\let\auto@bib@innerbib\@empty
\bibitem [{\citenamefont {Hen}\ \emph {et~al.}(2017)\citenamefont {Hen},
  \citenamefont {Miller}, \citenamefont {Piasetzky},\ and\ \citenamefont
  {Weinstein}}]{Hen:2016kwk}%
  \BibitemOpen
  \bibfield  {author} {\bibinfo {author} {\bibfnamefont {O.}~\bibnamefont
  {Hen}}, \bibinfo {author} {\bibfnamefont {G.~A.}\ \bibnamefont {Miller}},
  \bibinfo {author} {\bibfnamefont {E.}~\bibnamefont {Piasetzky}}, \ and\
  \bibinfo {author} {\bibfnamefont {L.~B.}\ \bibnamefont {Weinstein}},\ }\href
  {\doibase 10.1103/RevModPhys.89.045002} {\bibfield  {journal} {\bibinfo
  {journal} {Rev. Mod. Phys.}\ }\textbf {\bibinfo {volume} {89}},\ \bibinfo
  {pages} {045002} (\bibinfo {year} {2017})}\BibitemShut {NoStop}%
\bibitem [{\citenamefont {Ciofi~degli Atti}(2015)}]{Atti:2015eda}%
  \BibitemOpen
  \bibfield  {author} {\bibinfo {author} {\bibfnamefont {C.}~\bibnamefont
  {Ciofi~degli Atti}},\ }\href {\doibase 10.1016/j.physrep.2015.06.002}
  {\bibfield  {journal} {\bibinfo  {journal} {Phys. Rept.}\ }\textbf {\bibinfo
  {volume} {590}},\ \bibinfo {pages} {1} (\bibinfo {year} {2015})}\BibitemShut
  {NoStop}%
\bibitem [{\citenamefont {Piasetzky}\ \emph {et~al.}(2006)\citenamefont
  {Piasetzky}, \citenamefont {Sargsian}, \citenamefont {Frankfurt},
  \citenamefont {Strikman},\ and\ \citenamefont {Watson}}]{piasetzky06}%
  \BibitemOpen
  \bibfield  {author} {\bibinfo {author} {\bibfnamefont {E.}~\bibnamefont
  {Piasetzky}}, \bibinfo {author} {\bibfnamefont {M.}~\bibnamefont {Sargsian}},
  \bibinfo {author} {\bibfnamefont {L.}~\bibnamefont {Frankfurt}}, \bibinfo
  {author} {\bibfnamefont {M.}~\bibnamefont {Strikman}}, \ and\ \bibinfo
  {author} {\bibfnamefont {J.~W.}\ \bibnamefont {Watson}},\ }\href {\doibase
  10.1103/PhysRevLett.97.162504} {\bibfield  {journal} {\bibinfo  {journal}
  {Phys. Rev. Lett.}\ }\textbf {\bibinfo {volume} {97}},\ \bibinfo {pages}
  {162504} (\bibinfo {year} {2006})}\BibitemShut {NoStop}%
\bibitem [{\citenamefont {Subedi}\ \emph {et~al.}(2008)\citenamefont {Subedi}
  \emph {et~al.}}]{subedi08}%
  \BibitemOpen
  \bibfield  {author} {\bibinfo {author} {\bibfnamefont {R.}~\bibnamefont
  {Subedi}} \emph {et~al.},\ }\href {\doibase 10.1126/science.1156675}
  {\bibfield  {journal} {\bibinfo  {journal} {Science}\ }\textbf {\bibinfo
  {volume} {320}},\ \bibinfo {pages} {1476} (\bibinfo {year} {2008})},\ \Eprint
  {http://arxiv.org/abs/0908.1514} {arXiv:0908.1514 [nucl-ex]} \BibitemShut
  {NoStop}%
\bibitem [{\citenamefont {Alvioli}\ \emph {et~al.}(2008)\citenamefont
  {Alvioli}, \citenamefont {degli Atti},\ and\ \citenamefont
  {Morita}}]{alvioli08}%
  \BibitemOpen
  \bibfield  {author} {\bibinfo {author} {\bibfnamefont {M.}~\bibnamefont
  {Alvioli}}, \bibinfo {author} {\bibfnamefont {C.~C.}\ \bibnamefont {degli
  Atti}}, \ and\ \bibinfo {author} {\bibfnamefont {H.}~\bibnamefont {Morita}},\
  }\href@noop {} {\bibfield  {journal} {\bibinfo  {journal} {Phys. Rev. Lett.}\
  }\textbf {\bibinfo {volume} {100}},\ \bibinfo {eid} {162503} (\bibinfo {year}
  {2008})}\BibitemShut {NoStop}%
\bibitem [{\citenamefont {Hen}\ \emph {et~al.}(2014)\citenamefont {Hen} \emph
  {et~al.}}]{hen14}%
  \BibitemOpen
  \bibfield  {author} {\bibinfo {author} {\bibfnamefont {O.}~\bibnamefont
  {Hen}} \emph {et~al.},\ }\href {\doibase 10.1126/science.1256785} {\bibfield
  {journal} {\bibinfo  {journal} {Science}\ }\textbf {\bibinfo {volume}
  {346}},\ \bibinfo {pages} {614} (\bibinfo {year} {2014})},\ \Eprint
  {http://arxiv.org/abs/1412.0138} {arXiv:1412.0138 [nucl-ex]} \BibitemShut
  {NoStop}%
\bibitem [{\citenamefont {Korover}\ \emph {et~al.}(2014)\citenamefont
  {Korover}, \citenamefont {Muangma}, \citenamefont {Hen} \emph
  {et~al.}}]{korover14}%
  \BibitemOpen
  \bibfield  {author} {\bibinfo {author} {\bibfnamefont {I.}~\bibnamefont
  {Korover}}, \bibinfo {author} {\bibfnamefont {N.}~\bibnamefont {Muangma}},
  \bibinfo {author} {\bibfnamefont {O.}~\bibnamefont {Hen}},  \emph {et~al.},\
  }\href {\doibase 10.1103/PhysRevLett.113.022501} {\bibfield  {journal}
  {\bibinfo  {journal} {Phys. Rev. Lett.}\ }\textbf {\bibinfo {volume} {113}},\
  \bibinfo {pages} {022501} (\bibinfo {year} {2014})}\BibitemShut {NoStop}%
\bibitem [{\citenamefont {Weiss}\ \emph {et~al.}(2018)\citenamefont {Weiss},
  \citenamefont {Cruz-Torres}, \citenamefont {Barnea}, \citenamefont
  {Piasetzky},\ and\ \citenamefont {Hen}}]{Weiss:2016obx}%
  \BibitemOpen
  \bibfield  {author} {\bibinfo {author} {\bibfnamefont {R.}~\bibnamefont
  {Weiss}}, \bibinfo {author} {\bibfnamefont {R.}~\bibnamefont {Cruz-Torres}},
  \bibinfo {author} {\bibfnamefont {N.}~\bibnamefont {Barnea}}, \bibinfo
  {author} {\bibfnamefont {E.}~\bibnamefont {Piasetzky}}, \ and\ \bibinfo
  {author} {\bibfnamefont {O.}~\bibnamefont {Hen}},\ }\href@noop {} {\bibfield
  {journal} {\bibinfo  {journal} {Phys. Lett. B}\ }\textbf {\bibinfo {volume}
  {780}},\ \bibinfo {pages} {211} (\bibinfo {year} {2018})}\BibitemShut
  {NoStop}%
\bibitem [{\citenamefont {Duer}\ \emph {et~al.}(2019)\citenamefont {Duer} \emph
  {et~al.}}]{Duer:2018sxh}%
  \BibitemOpen
  \bibfield  {author} {\bibinfo {author} {\bibfnamefont {M.}~\bibnamefont
  {Duer}} \emph {et~al.} (\bibinfo {collaboration} {CLAS Collaboration}),\
  }\href {\doibase 10.1103/PhysRevLett.122.172502} {\bibfield  {journal}
  {\bibinfo  {journal} {Phys. Rev. Lett.}\ }\textbf {\bibinfo {volume} {122}},\
  \bibinfo {pages} {172502} (\bibinfo {year} {2019})},\ \Eprint
  {http://arxiv.org/abs/1810.05343} {arXiv:1810.05343 [nucl-ex]} \BibitemShut
  {NoStop}%
\bibitem [{\citenamefont {Korover}\ \emph {et~al.}(2020)\citenamefont {Korover}
  \emph {et~al.}}]{Korover:2020lqf}%
  \BibitemOpen
  \bibfield  {author} {\bibinfo {author} {\bibfnamefont {I.}~\bibnamefont
  {Korover}} \emph {et~al.} (\bibinfo {collaboration} {CLAS}),\ }\href@noop {}
  {\  (\bibinfo {year} {2020})},\ \Eprint {http://arxiv.org/abs/2004.07304}
  {arXiv:2004.07304 [nucl-ex]} \BibitemShut {NoStop}%
\bibitem [{\citenamefont {Tan}(2008{\natexlab{a}})}]{Tan08a}%
  \BibitemOpen
  \bibfield  {author} {\bibinfo {author} {\bibfnamefont {S.}~\bibnamefont
  {Tan}},\ }\href@noop {} {\bibfield  {journal} {\bibinfo  {journal} {Annals of
  Physics}\ }\textbf {\bibinfo {volume} {323}},\ \bibinfo {pages} {2952}
  (\bibinfo {year} {2008}{\natexlab{a}})}\BibitemShut {NoStop}%
\bibitem [{\citenamefont {Tan}(2008{\natexlab{b}})}]{Tan08b}%
  \BibitemOpen
  \bibfield  {author} {\bibinfo {author} {\bibfnamefont {S.}~\bibnamefont
  {Tan}},\ }\href {\doibase http://dx.doi.org/10.1016/j.aop.2008.03.005}
  {\bibfield  {journal} {\bibinfo  {journal} {Annals of Physics}\ }\textbf
  {\bibinfo {volume} {323}},\ \bibinfo {pages} {2971} (\bibinfo {year}
  {2008}{\natexlab{b}})}\BibitemShut {NoStop}%
\bibitem [{\citenamefont {Tan}(2008{\natexlab{c}})}]{Tan08c}%
  \BibitemOpen
  \bibfield  {author} {\bibinfo {author} {\bibfnamefont {S.}~\bibnamefont
  {Tan}},\ }\href {\doibase http://dx.doi.org/10.1016/j.aop.2008.03.003}
  {\bibfield  {journal} {\bibinfo  {journal} {Annals of Physics}\ }\textbf
  {\bibinfo {volume} {323}},\ \bibinfo {pages} {2987} (\bibinfo {year}
  {2008}{\natexlab{c}})}\BibitemShut {NoStop}%
\bibitem [{\citenamefont {Braaten}(2012)}]{Braaten12}%
  \BibitemOpen
  \bibfield  {author} {\bibinfo {author} {\bibfnamefont {E.}~\bibnamefont
  {Braaten}},\ }in\ \href@noop {} {\emph {\bibinfo {booktitle} {The BCS-BEC
  Crossover and the Unitary Fermi Gas}}},\ \bibinfo {editor} {edited by\
  \bibinfo {editor} {\bibfnamefont {W.}~\bibnamefont {Zwerger}}}\ (\bibinfo
  {publisher} {Springer},\ \bibinfo {address} {Berlin},\ \bibinfo {year}
  {2012})\BibitemShut {NoStop}%
\bibitem [{\citenamefont {Weiss}\ \emph
  {et~al.}(2015{\natexlab{a}})\citenamefont {Weiss}, \citenamefont {Bazak},\
  and\ \citenamefont {Barnea}}]{Weiss14}%
  \BibitemOpen
  \bibfield  {author} {\bibinfo {author} {\bibfnamefont {R.}~\bibnamefont
  {Weiss}}, \bibinfo {author} {\bibfnamefont {B.}~\bibnamefont {Bazak}}, \ and\
  \bibinfo {author} {\bibfnamefont {N.}~\bibnamefont {Barnea}},\ }\href
  {\doibase 10.1103/PhysRevLett.114.012501} {\bibfield  {journal} {\bibinfo
  {journal} {Phys. Rev. Lett.}\ }\textbf {\bibinfo {volume} {114}},\ \bibinfo
  {pages} {012501} (\bibinfo {year} {2015}{\natexlab{a}})}\BibitemShut
  {NoStop}%
\bibitem [{\citenamefont {Weiss}\ \emph
  {et~al.}(2015{\natexlab{b}})\citenamefont {Weiss}, \citenamefont {Bazak},\
  and\ \citenamefont {Barnea}}]{Weiss:2015mba}%
  \BibitemOpen
  \bibfield  {author} {\bibinfo {author} {\bibfnamefont {R.}~\bibnamefont
  {Weiss}}, \bibinfo {author} {\bibfnamefont {B.}~\bibnamefont {Bazak}}, \ and\
  \bibinfo {author} {\bibfnamefont {N.}~\bibnamefont {Barnea}},\ }\href
  {\doibase 10.1103/PhysRevC.92.054311} {\bibfield  {journal} {\bibinfo
  {journal} {Phys. Rev.}\ }\textbf {\bibinfo {volume} {C92}},\ \bibinfo {pages}
  {054311} (\bibinfo {year} {2015}{\natexlab{b}})},\ \Eprint
  {http://arxiv.org/abs/1503.07047} {arXiv:1503.07047 [nucl-th]} \BibitemShut
  {NoStop}%
\bibitem [{\citenamefont {Cruz-Torres}\ \emph {et~al.}(2018)\citenamefont
  {Cruz-Torres}, \citenamefont {Schmidt}, \citenamefont {Miller}, \citenamefont
  {Weinstein}, \citenamefont {Barnea}, \citenamefont {Weiss}, \citenamefont
  {Piasetzky},\ and\ \citenamefont {Hen}}]{Cruz-Torres:2017sjy}%
  \BibitemOpen
  \bibfield  {author} {\bibinfo {author} {\bibfnamefont {R.}~\bibnamefont
  {Cruz-Torres}}, \bibinfo {author} {\bibfnamefont {A.}~\bibnamefont
  {Schmidt}}, \bibinfo {author} {\bibfnamefont {G.~A.}\ \bibnamefont {Miller}},
  \bibinfo {author} {\bibfnamefont {L.~B.}\ \bibnamefont {Weinstein}}, \bibinfo
  {author} {\bibfnamefont {N.}~\bibnamefont {Barnea}}, \bibinfo {author}
  {\bibfnamefont {R.}~\bibnamefont {Weiss}}, \bibinfo {author} {\bibfnamefont
  {E.}~\bibnamefont {Piasetzky}}, \ and\ \bibinfo {author} {\bibfnamefont
  {O.}~\bibnamefont {Hen}},\ }\href {\doibase 10.1016/j.physletb.2018.07.069}
  {\bibfield  {journal} {\bibinfo  {journal} {Phys. Lett.}\ }\textbf {\bibinfo
  {volume} {B785}},\ \bibinfo {pages} {304} (\bibinfo {year} {2018})},\ \Eprint
  {http://arxiv.org/abs/1710.07966} {arXiv:1710.07966 [nucl-th]} \BibitemShut
  {NoStop}%
\bibitem [{\citenamefont {Cruz-Torres}\ \emph {et~al.}(2020)\citenamefont
  {Cruz-Torres}, \citenamefont {Lonardoni}, \citenamefont {Weiss},
  \citenamefont {Barnea}, \citenamefont {Higinbotham}, \citenamefont
  {Piasetzky}, \citenamefont {Schmidt}, \citenamefont {Weinstein},
  \citenamefont {Wiringa},\ and\ \citenamefont {Hen}}]{Cruz-Torres:2019fum}%
  \BibitemOpen
  \bibfield  {author} {\bibinfo {author} {\bibfnamefont {R.}~\bibnamefont
  {Cruz-Torres}}, \bibinfo {author} {\bibfnamefont {D.}~\bibnamefont
  {Lonardoni}}, \bibinfo {author} {\bibfnamefont {R.}~\bibnamefont {Weiss}},
  \bibinfo {author} {\bibfnamefont {N.}~\bibnamefont {Barnea}}, \bibinfo
  {author} {\bibfnamefont {D.~W.}\ \bibnamefont {Higinbotham}}, \bibinfo
  {author} {\bibfnamefont {E.}~\bibnamefont {Piasetzky}}, \bibinfo {author}
  {\bibfnamefont {A.}~\bibnamefont {Schmidt}}, \bibinfo {author} {\bibfnamefont
  {L.~B.}\ \bibnamefont {Weinstein}}, \bibinfo {author} {\bibfnamefont {R.~B.}\
  \bibnamefont {Wiringa}}, \ and\ \bibinfo {author} {\bibfnamefont
  {O.}~\bibnamefont {Hen}},\ }\href@noop {} {\bibfield  {journal} {\bibinfo
  {journal} {Nature Physics}\ } (\bibinfo {year} {2020})},\ \Eprint
  {http://arxiv.org/abs/1907.03658} {arXiv:1907.03658 [nucl-th]} \BibitemShut
  {NoStop}%
\bibitem [{\citenamefont {Weiss}\ \emph {et~al.}(2019)\citenamefont {Weiss},
  \citenamefont {Korover}, \citenamefont {Piasetzky}, \citenamefont {Hen},\
  and\ \citenamefont {Barnea}}]{Weiss:2018tbu}%
  \BibitemOpen
  \bibfield  {author} {\bibinfo {author} {\bibfnamefont {R.}~\bibnamefont
  {Weiss}}, \bibinfo {author} {\bibfnamefont {I.}~\bibnamefont {Korover}},
  \bibinfo {author} {\bibfnamefont {E.}~\bibnamefont {Piasetzky}}, \bibinfo
  {author} {\bibfnamefont {O.}~\bibnamefont {Hen}}, \ and\ \bibinfo {author}
  {\bibfnamefont {N.}~\bibnamefont {Barnea}},\ }\href {\doibase
  10.1016/j.physletb.2019.02.019} {\bibfield  {journal} {\bibinfo  {journal}
  {Phys. Lett.}\ }\textbf {\bibinfo {volume} {B791}},\ \bibinfo {pages} {242}
  (\bibinfo {year} {2019})},\ \Eprint {http://arxiv.org/abs/1806.10217}
  {arXiv:1806.10217 [nucl-th]} \BibitemShut {NoStop}%
\bibitem [{\citenamefont {Schmidt}\ \emph {et~al.}(2020)\citenamefont {Schmidt}
  \emph {et~al.}}]{schmidt20}%
  \BibitemOpen
  \bibfield  {author} {\bibinfo {author} {\bibfnamefont {A.}~\bibnamefont
  {Schmidt}} \emph {et~al.} (\bibinfo {collaboration} {CLAS}),\ }\href
  {\doibase 10.1038/s41586-020-2021-6} {\bibfield  {journal} {\bibinfo
  {journal} {Nature}\ }\textbf {\bibinfo {volume} {578}},\ \bibinfo {pages}
  {540} (\bibinfo {year} {2020})},\ \Eprint {http://arxiv.org/abs/2004.11221}
  {arXiv:2004.11221 [nucl-ex]} \BibitemShut {NoStop}%
\bibitem [{\citenamefont {Hen}\ \emph {et~al.}(2015)\citenamefont {Hen},
  \citenamefont {Li}, \citenamefont {Guo}, \citenamefont {Weinstein},\ and\
  \citenamefont {Piasetzky}}]{hen15}%
  \BibitemOpen
  \bibfield  {author} {\bibinfo {author} {\bibfnamefont {O.}~\bibnamefont
  {Hen}}, \bibinfo {author} {\bibfnamefont {B.-A.}\ \bibnamefont {Li}},
  \bibinfo {author} {\bibfnamefont {W.-J.}\ \bibnamefont {Guo}}, \bibinfo
  {author} {\bibfnamefont {L.~B.}\ \bibnamefont {Weinstein}}, \ and\ \bibinfo
  {author} {\bibfnamefont {E.}~\bibnamefont {Piasetzky}},\ }\href {\doibase
  10.1103/PhysRevC.91.025803} {\bibfield  {journal} {\bibinfo  {journal} {Phys.
  Rev. C}\ }\textbf {\bibinfo {volume} {91}},\ \bibinfo {pages} {025803}
  (\bibinfo {year} {2015})}\BibitemShut {NoStop}%
\bibitem [{\citenamefont {Cai}\ and\ \citenamefont {Li}(2016)}]{Cai:2015xga}%
  \BibitemOpen
  \bibfield  {author} {\bibinfo {author} {\bibfnamefont {B.-J.}\ \bibnamefont
  {Cai}}\ and\ \bibinfo {author} {\bibfnamefont {B.-A.}\ \bibnamefont {Li}},\
  }\href {\doibase 10.1103/PhysRevC.93.014619} {\bibfield  {journal} {\bibinfo
  {journal} {Phys. Rev.}\ }\textbf {\bibinfo {volume} {C93}},\ \bibinfo {pages}
  {014619} (\bibinfo {year} {2016})}\BibitemShut {NoStop}%
\bibitem [{\citenamefont {Li}\ \emph {et~al.}(2018)\citenamefont {Li},
  \citenamefont {Cai}, \citenamefont {Chen},\ and\ \citenamefont
  {Xu}}]{Li:2018lpy}%
  \BibitemOpen
  \bibfield  {author} {\bibinfo {author} {\bibfnamefont {B.-A.}\ \bibnamefont
  {Li}}, \bibinfo {author} {\bibfnamefont {B.-J.}\ \bibnamefont {Cai}},
  \bibinfo {author} {\bibfnamefont {L.-W.}\ \bibnamefont {Chen}}, \ and\
  \bibinfo {author} {\bibfnamefont {J.}~\bibnamefont {Xu}},\ }\href {\doibase
  10.1016/j.ppnp.2018.01.001} {\bibfield  {journal} {\bibinfo  {journal} {Prog.
  Part. Nucl. Phys.}\ }\textbf {\bibinfo {volume} {99}},\ \bibinfo {pages} {29}
  (\bibinfo {year} {2018})},\ \Eprint {http://arxiv.org/abs/1801.01213}
  {arXiv:1801.01213 [nucl-th]} \BibitemShut {NoStop}%
\bibitem [{\citenamefont {Souza}\ \emph {et~al.}(2020)\citenamefont {Souza},
  \citenamefont {Negreiros}, \citenamefont {Dutra}, \citenamefont {Menezes},\
  and\ \citenamefont {Louren\c{c}o}}]{Souza:2020gjs}%
  \BibitemOpen
  \bibfield  {author} {\bibinfo {author} {\bibfnamefont {L.~A.}\ \bibnamefont
  {Souza}}, \bibinfo {author} {\bibfnamefont {R.}~\bibnamefont {Negreiros}},
  \bibinfo {author} {\bibfnamefont {M.}~\bibnamefont {Dutra}}, \bibinfo
  {author} {\bibfnamefont {D.~P.}\ \bibnamefont {Menezes}}, \ and\ \bibinfo
  {author} {\bibfnamefont {O.}~\bibnamefont {Louren\c{c}o}},\ }\href@noop {} {\
   (\bibinfo {year} {2020})},\ \Eprint {http://arxiv.org/abs/2004.10309}
  {arXiv:2004.10309 [nucl-th]} \BibitemShut {NoStop}%
\bibitem [{\citenamefont {Weinstein}\ \emph {et~al.}(2011)\citenamefont
  {Weinstein}, \citenamefont {Piasetzky}, \citenamefont {Higinbotham},
  \citenamefont {Gomez}, \citenamefont {Hen},\ and\ \citenamefont
  {Shneor}}]{weinstein11}%
  \BibitemOpen
  \bibfield  {author} {\bibinfo {author} {\bibfnamefont {L.~B.}\ \bibnamefont
  {Weinstein}}, \bibinfo {author} {\bibfnamefont {E.}~\bibnamefont
  {Piasetzky}}, \bibinfo {author} {\bibfnamefont {D.~W.}\ \bibnamefont
  {Higinbotham}}, \bibinfo {author} {\bibfnamefont {J.}~\bibnamefont {Gomez}},
  \bibinfo {author} {\bibfnamefont {O.}~\bibnamefont {Hen}}, \ and\ \bibinfo
  {author} {\bibfnamefont {R.}~\bibnamefont {Shneor}},\ }\href {\doibase
  10.1103/PhysRevLett.106.052301} {\bibfield  {journal} {\bibinfo  {journal}
  {Phys. Rev. Lett.}\ }\textbf {\bibinfo {volume} {106}},\ \bibinfo {pages}
  {052301} (\bibinfo {year} {2011})}\BibitemShut {NoStop}%
\bibitem [{\citenamefont {Hen}\ \emph {et~al.}(2012)\citenamefont {Hen},
  \citenamefont {Piasetzky},\ and\ \citenamefont {Weinstein}}]{Hen12}%
  \BibitemOpen
  \bibfield  {author} {\bibinfo {author} {\bibfnamefont {O.}~\bibnamefont
  {Hen}}, \bibinfo {author} {\bibfnamefont {E.}~\bibnamefont {Piasetzky}}, \
  and\ \bibinfo {author} {\bibfnamefont {L.~B.}\ \bibnamefont {Weinstein}},\
  }\href {\doibase 10.1103/PhysRevC.85.047301} {\bibfield  {journal} {\bibinfo
  {journal} {Phys. Rev. C}\ }\textbf {\bibinfo {volume} {85}},\ \bibinfo
  {pages} {047301} (\bibinfo {year} {2012})}\BibitemShut {NoStop}%
\bibitem [{\citenamefont {Hen}\ \emph {et~al.}(2013)\citenamefont {Hen},
  \citenamefont {Higinbotham}, \citenamefont {Miller}, \citenamefont
  {Piasetzky},\ and\ \citenamefont {Weinstein}}]{Hen:2013oha}%
  \BibitemOpen
  \bibfield  {author} {\bibinfo {author} {\bibfnamefont {O.}~\bibnamefont
  {Hen}}, \bibinfo {author} {\bibfnamefont {D.~W.}\ \bibnamefont
  {Higinbotham}}, \bibinfo {author} {\bibfnamefont {G.~A.}\ \bibnamefont
  {Miller}}, \bibinfo {author} {\bibfnamefont {E.}~\bibnamefont {Piasetzky}}, \
  and\ \bibinfo {author} {\bibfnamefont {L.~B.}\ \bibnamefont {Weinstein}},\
  }\href {\doibase 10.1142/S0218301313300178} {\bibfield  {journal} {\bibinfo
  {journal} {Int. J. Mod. Phys.}\ }\textbf {\bibinfo {volume} {E22}},\ \bibinfo
  {pages} {1330017} (\bibinfo {year} {2013})},\ \Eprint
  {http://arxiv.org/abs/1304.2813} {arXiv:1304.2813 [nucl-th]} \BibitemShut
  {NoStop}%
\bibitem [{\citenamefont {Chen}\ \emph {et~al.}(2017)\citenamefont {Chen},
  \citenamefont {Detmold}, \citenamefont {Lynn},\ and\ \citenamefont
  {Schwenk}}]{Chen:2016bde}%
  \BibitemOpen
  \bibfield  {author} {\bibinfo {author} {\bibfnamefont {J.-W.}\ \bibnamefont
  {Chen}}, \bibinfo {author} {\bibfnamefont {W.}~\bibnamefont {Detmold}},
  \bibinfo {author} {\bibfnamefont {J.~E.}\ \bibnamefont {Lynn}}, \ and\
  \bibinfo {author} {\bibfnamefont {A.}~\bibnamefont {Schwenk}},\ }\href
  {\doibase 10.1103/PhysRevLett.119.262502} {\bibfield  {journal} {\bibinfo
  {journal} {Phys. Rev. Lett.}\ }\textbf {\bibinfo {volume} {119}},\ \bibinfo
  {pages} {262502} (\bibinfo {year} {2017})},\ \Eprint
  {http://arxiv.org/abs/1607.03065} {arXiv:1607.03065 [hep-ph]} \BibitemShut
  {NoStop}%
\bibitem [{\citenamefont {{C. Ciofi degli Atti, L.L. Frankfurt, L.P. Kaptari
  and M.I. Strikman}}(2007)}]{Ciofi07}%
  \BibitemOpen
  \bibfield  {author} {\bibinfo {author} {\bibnamefont {{C. Ciofi degli Atti,
  L.L. Frankfurt, L.P. Kaptari and M.I. Strikman}}},\ }\href@noop {} {\bibfield
   {journal} {\bibinfo  {journal} {Phys. Rev. C}\ }\textbf {\bibinfo {volume}
  {76}},\ \bibinfo {pages} {055206} (\bibinfo {year} {2007})}\BibitemShut
  {NoStop}%
\bibitem [{\citenamefont {Schmookler}\ \emph {et~al.}(2019)\citenamefont
  {Schmookler} \emph {et~al.}}]{Schmookler:2019nvf}%
  \BibitemOpen
  \bibfield  {author} {\bibinfo {author} {\bibfnamefont {B.}~\bibnamefont
  {Schmookler}} \emph {et~al.} (\bibinfo {collaboration} {CLAS
  Collaboration}),\ }\href {\doibase 10.1038/s41586-019-0925-9} {\bibfield
  {journal} {\bibinfo  {journal} {Nature}\ }\textbf {\bibinfo {volume} {566}},\
  \bibinfo {pages} {354} (\bibinfo {year} {2019})}\BibitemShut {NoStop}%
\bibitem [{\citenamefont {Arrington}\ and\ \citenamefont
  {Fomin}(2019)}]{Arrington:2019wky}%
  \BibitemOpen
  \bibfield  {author} {\bibinfo {author} {\bibfnamefont {J.}~\bibnamefont
  {Arrington}}\ and\ \bibinfo {author} {\bibfnamefont {N.}~\bibnamefont
  {Fomin}},\ }\href {\doibase 10.1103/PhysRevLett.123.042501} {\bibfield
  {journal} {\bibinfo  {journal} {Phys. Rev. Lett.}\ }\textbf {\bibinfo
  {volume} {123}},\ \bibinfo {pages} {042501} (\bibinfo {year} {2019})},\
  \Eprint {http://arxiv.org/abs/1903.12535} {arXiv:1903.12535 [nucl-ex]}
  \BibitemShut {NoStop}%
\bibitem [{\citenamefont {Hen}\ \emph {et~al.}(2019)\citenamefont {Hen},
  \citenamefont {Hauenstein}, \citenamefont {Higinbotham}, \citenamefont
  {Miller}, \citenamefont {Piasetzky}, \citenamefont {Schmidt}, \citenamefont
  {Segarra}, \citenamefont {Strikman},\ and\ \citenamefont
  {Weinstein}}]{Hen:2019jzn}%
  \BibitemOpen
  \bibfield  {author} {\bibinfo {author} {\bibfnamefont {O.}~\bibnamefont
  {Hen}}, \bibinfo {author} {\bibfnamefont {F.}~\bibnamefont {Hauenstein}},
  \bibinfo {author} {\bibfnamefont {D.~W.}\ \bibnamefont {Higinbotham}},
  \bibinfo {author} {\bibfnamefont {G.~A.}\ \bibnamefont {Miller}}, \bibinfo
  {author} {\bibfnamefont {E.}~\bibnamefont {Piasetzky}}, \bibinfo {author}
  {\bibfnamefont {A.}~\bibnamefont {Schmidt}}, \bibinfo {author} {\bibfnamefont
  {E.~P.}\ \bibnamefont {Segarra}}, \bibinfo {author} {\bibfnamefont
  {M.}~\bibnamefont {Strikman}}, \ and\ \bibinfo {author} {\bibfnamefont
  {L.~B.}\ \bibnamefont {Weinstein}},\ }\href@noop {} {\  (\bibinfo {year}
  {2019})},\ \Eprint {http://arxiv.org/abs/1905.02172} {arXiv:1905.02172
  [nucl-ex]} \BibitemShut {NoStop}%
\bibitem [{\citenamefont {Hen}\ \emph {et~al.}(2011)\citenamefont {Hen},
  \citenamefont {Accardi}, \citenamefont {Melnitchouk},\ and\ \citenamefont
  {Piasetzky}}]{hen11}%
  \BibitemOpen
  \bibfield  {author} {\bibinfo {author} {\bibfnamefont {O.}~\bibnamefont
  {Hen}}, \bibinfo {author} {\bibfnamefont {A.}~\bibnamefont {Accardi}},
  \bibinfo {author} {\bibfnamefont {W.}~\bibnamefont {Melnitchouk}}, \ and\
  \bibinfo {author} {\bibfnamefont {E.}~\bibnamefont {Piasetzky}},\ }\href
  {\doibase 10.1103/PhysRevD.84.117501} {\bibfield  {journal} {\bibinfo
  {journal} {Phys. Rev. D}\ }\textbf {\bibinfo {volume} {84}},\ \bibinfo
  {pages} {117501} (\bibinfo {year} {2011})}\BibitemShut {NoStop}%
\bibitem [{\citenamefont {Segarra}\ \emph {et~al.}(2020)\citenamefont
  {Segarra}, \citenamefont {Schmidt}, \citenamefont {Kutz}, \citenamefont
  {Higinbotham}, \citenamefont {Piasetzky}, \citenamefont {Strikman},
  \citenamefont {Weinstein},\ and\ \citenamefont {Hen}}]{Segarra:2019gbp}%
  \BibitemOpen
  \bibfield  {author} {\bibinfo {author} {\bibfnamefont {E.~P.}\ \bibnamefont
  {Segarra}}, \bibinfo {author} {\bibfnamefont {A.}~\bibnamefont {Schmidt}},
  \bibinfo {author} {\bibfnamefont {T.}~\bibnamefont {Kutz}}, \bibinfo {author}
  {\bibfnamefont {D.~W.}\ \bibnamefont {Higinbotham}}, \bibinfo {author}
  {\bibfnamefont {E.}~\bibnamefont {Piasetzky}}, \bibinfo {author}
  {\bibfnamefont {M.}~\bibnamefont {Strikman}}, \bibinfo {author}
  {\bibfnamefont {L.~B.}\ \bibnamefont {Weinstein}}, \ and\ \bibinfo {author}
  {\bibfnamefont {O.}~\bibnamefont {Hen}},\ }\href {\doibase
  10.1103/PhysRevLett.124.092002} {\bibfield  {journal} {\bibinfo  {journal}
  {Phys. Rev. Lett.}\ }\textbf {\bibinfo {volume} {124}},\ \bibinfo {pages}
  {092002} (\bibinfo {year} {2020})}\BibitemShut {NoStop}%
\bibitem [{\citenamefont {Frankfurt}\ \emph {et~al.}(1993)\citenamefont
  {Frankfurt}, \citenamefont {Strikman}, \citenamefont {Day},\ and\
  \citenamefont {Sargsyan}}]{frankfurt93}%
  \BibitemOpen
  \bibfield  {author} {\bibinfo {author} {\bibfnamefont {L.}~\bibnamefont
  {Frankfurt}}, \bibinfo {author} {\bibfnamefont {M.}~\bibnamefont {Strikman}},
  \bibinfo {author} {\bibfnamefont {D.}~\bibnamefont {Day}}, \ and\ \bibinfo
  {author} {\bibfnamefont {M.}~\bibnamefont {Sargsyan}},\ }\href@noop {}
  {\bibfield  {journal} {\bibinfo  {journal} {Phys. Rev. C}\ }\textbf {\bibinfo
  {volume} {48}},\ \bibinfo {pages} {2451} (\bibinfo {year}
  {1993})}\BibitemShut {NoStop}%
\bibitem [{\citenamefont {Egiyan}\ \emph {et~al.}(2003)\citenamefont {Egiyan}
  \emph {et~al.}}]{egiyan02}%
  \BibitemOpen
  \bibfield  {author} {\bibinfo {author} {\bibfnamefont {K.}~\bibnamefont
  {Egiyan}} \emph {et~al.} (\bibinfo {collaboration} {CLAS Collaboration}),\
  }\href@noop {} {\bibfield  {journal} {\bibinfo  {journal} {Phys. Rev. C}\
  }\textbf {\bibinfo {volume} {68}},\ \bibinfo {pages} {014313} (\bibinfo
  {year} {2003})}\BibitemShut {NoStop}%
\bibitem [{\citenamefont {Egiyan}\ \emph {et~al.}(2006)\citenamefont {Egiyan}
  \emph {et~al.}}]{egiyan06}%
  \BibitemOpen
  \bibfield  {author} {\bibinfo {author} {\bibfnamefont {K.}~\bibnamefont
  {Egiyan}} \emph {et~al.} (\bibinfo {collaboration} {CLAS Collaboration}),\
  }\href@noop {} {\bibfield  {journal} {\bibinfo  {journal} {Phys. Rev. Lett.}\
  }\textbf {\bibinfo {volume} {96}},\ \bibinfo {pages} {082501} (\bibinfo
  {year} {2006})}\BibitemShut {NoStop}%
\bibitem [{\citenamefont {Fomin}\ \emph {et~al.}(2012)\citenamefont {Fomin}
  \emph {et~al.}}]{fomin12}%
  \BibitemOpen
  \bibfield  {author} {\bibinfo {author} {\bibfnamefont {N.}~\bibnamefont
  {Fomin}} \emph {et~al.},\ }\href@noop {} {\bibfield  {journal} {\bibinfo
  {journal} {Phys. Rev. Lett.}\ }\textbf {\bibinfo {volume} {108}},\ \bibinfo
  {pages} {092502} (\bibinfo {year} {2012})}\BibitemShut {NoStop}%
\bibitem [{\citenamefont {Nguyen}\ \emph {et~al.}(2020)\citenamefont {Nguyen}
  \emph {et~al.}}]{Nguyen:2020mgo}%
  \BibitemOpen
  \bibfield  {author} {\bibinfo {author} {\bibfnamefont {D.}~\bibnamefont
  {Nguyen}} \emph {et~al.},\ }\href@noop {} {\  (\bibinfo {year} {2020})},\
  \Eprint {http://arxiv.org/abs/2004.11448} {arXiv:2004.11448 [nucl-ex]}
  \BibitemShut {NoStop}%
\bibitem [{\citenamefont {Cohen}\ \emph {et~al.}(2018)\citenamefont {Cohen}
  \emph {et~al.}}]{Cohen:2018gzh}%
  \BibitemOpen
  \bibfield  {author} {\bibinfo {author} {\bibfnamefont {E.~O.}\ \bibnamefont
  {Cohen}} \emph {et~al.} (\bibinfo {collaboration} {CLAS Collaboration}),\
  }\href {\doibase 10.1103/PhysRevLett.121.092501} {\bibfield  {journal}
  {\bibinfo  {journal} {Phys. Rev. Lett.}\ }\textbf {\bibinfo {volume} {121}},\
  \bibinfo {pages} {092501} (\bibinfo {year} {2018})},\ \Eprint
  {http://arxiv.org/abs/1805.01981} {arXiv:1805.01981 [nucl-ex]} \BibitemShut
  {NoStop}%
\bibitem [{\citenamefont {Frankfurt}\ and\ \citenamefont
  {Strikman}(1981)}]{Frankfurt81}%
  \BibitemOpen
  \bibfield  {author} {\bibinfo {author} {\bibfnamefont {L.~L.}\ \bibnamefont
  {Frankfurt}}\ and\ \bibinfo {author} {\bibfnamefont {M.~I.}\ \bibnamefont
  {Strikman}},\ }\href@noop {} {\bibfield  {journal} {\bibinfo  {journal}
  {Phys. Rep.}\ }\textbf {\bibinfo {volume} {76}},\ \bibinfo {pages} {215}
  (\bibinfo {year} {1981})}\BibitemShut {NoStop}%
\bibitem [{\citenamefont {Frankfurt}\ and\ \citenamefont
  {Strikman}(1988)}]{Frankfurt88}%
  \BibitemOpen
  \bibfield  {author} {\bibinfo {author} {\bibfnamefont {L.}~\bibnamefont
  {Frankfurt}}\ and\ \bibinfo {author} {\bibfnamefont {M.}~\bibnamefont
  {Strikman}},\ }\href@noop {} {\bibfield  {journal} {\bibinfo  {journal}
  {Phys. Rep.}\ }\textbf {\bibinfo {volume} {160}},\ \bibinfo {pages} {235 }
  (\bibinfo {year} {1988})}\BibitemShut {NoStop}%
\bibitem [{\citenamefont {Ciofi~degli Atti}\ and\ \citenamefont
  {Simula}(1994)}]{CiofidegliAtti:1994ys}%
  \BibitemOpen
  \bibfield  {author} {\bibinfo {author} {\bibfnamefont {C.}~\bibnamefont
  {Ciofi~degli Atti}}\ and\ \bibinfo {author} {\bibfnamefont {S.}~\bibnamefont
  {Simula}},\ }\href {\doibase 10.1016/0370-2693(94)90010-8} {\bibfield
  {journal} {\bibinfo  {journal} {Phys.\ Lett.\ B}\ }\textbf {\bibinfo {volume}
  {325}},\ \bibinfo {pages} {276} (\bibinfo {year} {1994})},\ \Eprint
  {http://arxiv.org/abs/nucl-th/9403001} {arXiv:nucl-th/9403001} \BibitemShut
  {NoStop}%
\bibitem [{\citenamefont {Ciofi~degli Atti}\ and\ \citenamefont
  {Simula}(1996)}]{CiofidegliAtti:1995qe}%
  \BibitemOpen
  \bibfield  {author} {\bibinfo {author} {\bibfnamefont {C.}~\bibnamefont
  {Ciofi~degli Atti}}\ and\ \bibinfo {author} {\bibfnamefont {S.}~\bibnamefont
  {Simula}},\ }\href {\doibase 10.1103/PhysRevC.53.1689} {\bibfield  {journal}
  {\bibinfo  {journal} {Phys. Rev. C}\ }\textbf {\bibinfo {volume} {53}},\
  \bibinfo {pages} {1689} (\bibinfo {year} {1996})}\BibitemShut {NoStop}%
\bibitem [{\citenamefont {Benhar}\ \emph {et~al.}(1995)\citenamefont {Benhar},
  \citenamefont {Fabrocini}, \citenamefont {Fantoni},\ and\ \citenamefont
  {Sick}}]{Benhar95}%
  \BibitemOpen
  \bibfield  {author} {\bibinfo {author} {\bibfnamefont {O.}~\bibnamefont
  {Benhar}}, \bibinfo {author} {\bibfnamefont {A.}~\bibnamefont {Fabrocini}},
  \bibinfo {author} {\bibfnamefont {S.}~\bibnamefont {Fantoni}}, \ and\
  \bibinfo {author} {\bibfnamefont {I.}~\bibnamefont {Sick}},\ }\href@noop {}
  {\bibfield  {journal} {\bibinfo  {journal} {Phys. Lett. B}\ }\textbf
  {\bibinfo {volume} {47}},\ \bibinfo {pages} {343} (\bibinfo {year}
  {1995})}\BibitemShut {NoStop}%
\bibitem [{\citenamefont {Frankfurt}\ \emph {et~al.}(2008)\citenamefont
  {Frankfurt}, \citenamefont {Sargsian},\ and\ \citenamefont
  {Strikman}}]{frankfurt08b}%
  \BibitemOpen
  \bibfield  {author} {\bibinfo {author} {\bibfnamefont {L.}~\bibnamefont
  {Frankfurt}}, \bibinfo {author} {\bibfnamefont {M.}~\bibnamefont {Sargsian}},
  \ and\ \bibinfo {author} {\bibfnamefont {M.}~\bibnamefont {Strikman}},\
  }\href {\doibase 10.1142/S0217751X08041207} {\bibfield  {journal} {\bibinfo
  {journal} {Int. J. Mod. Phys. A}\ }\textbf {\bibinfo {volume} {23}},\
  \bibinfo {pages} {2991} (\bibinfo {year} {2008})},\ \Eprint
  {http://arxiv.org/abs/0806.4412} {arXiv:0806.4412 [nucl-th]} \BibitemShut
  {NoStop}%
\bibitem [{\citenamefont {Fomin}\ \emph {et~al.}(2017)\citenamefont {Fomin},
  \citenamefont {Higinbotham}, \citenamefont {Sargsian},\ and\ \citenamefont
  {Solvignon}}]{Fomin:2017ydn}%
  \BibitemOpen
  \bibfield  {author} {\bibinfo {author} {\bibfnamefont {N.}~\bibnamefont
  {Fomin}}, \bibinfo {author} {\bibfnamefont {D.}~\bibnamefont {Higinbotham}},
  \bibinfo {author} {\bibfnamefont {M.}~\bibnamefont {Sargsian}}, \ and\
  \bibinfo {author} {\bibfnamefont {P.}~\bibnamefont {Solvignon}},\ }\href
  {\doibase 10.1146/annurev-nucl-102115-044939} {\bibfield  {journal} {\bibinfo
   {journal} {Ann.\ Rev.\ Nucl.\ Part.\ Sci.}\ }\textbf {\bibinfo {volume}
  {67}},\ \bibinfo {pages} {129} (\bibinfo {year} {2017})},\ \Eprint
  {http://arxiv.org/abs/1708.08581} {arXiv:1708.08581 [nucl-th]} \BibitemShut
  {NoStop}%
\bibitem [{\citenamefont {Schiavilla}\ \emph {et~al.}(2007)\citenamefont
  {Schiavilla}, \citenamefont {Wiringa}, \citenamefont {Pieper},\ and\
  \citenamefont {Carlson}}]{schiavilla07}%
  \BibitemOpen
  \bibfield  {author} {\bibinfo {author} {\bibfnamefont {R.}~\bibnamefont
  {Schiavilla}}, \bibinfo {author} {\bibfnamefont {R.~B.}\ \bibnamefont
  {Wiringa}}, \bibinfo {author} {\bibfnamefont {S.~C.}\ \bibnamefont {Pieper}},
  \ and\ \bibinfo {author} {\bibfnamefont {J.}~\bibnamefont {Carlson}},\
  }\href@noop {} {\bibfield  {journal} {\bibinfo  {journal} {Phys. Rev. Lett.}\
  }\textbf {\bibinfo {volume} {98}},\ \bibinfo {eid} {132501} (\bibinfo {year}
  {2007})}\BibitemShut {NoStop}%
\bibitem [{\citenamefont {Feldmeier}\ \emph {et~al.}(2011)\citenamefont
  {Feldmeier}, \citenamefont {Horiuchi}, \citenamefont {Neff},\ and\
  \citenamefont {Suzuki}}]{Feldmeier:2011qy}%
  \BibitemOpen
  \bibfield  {author} {\bibinfo {author} {\bibfnamefont {H.}~\bibnamefont
  {Feldmeier}}, \bibinfo {author} {\bibfnamefont {W.}~\bibnamefont {Horiuchi}},
  \bibinfo {author} {\bibfnamefont {T.}~\bibnamefont {Neff}}, \ and\ \bibinfo
  {author} {\bibfnamefont {Y.}~\bibnamefont {Suzuki}},\ }\href {\doibase
  10.1103/PhysRevC.84.054003} {\bibfield  {journal} {\bibinfo  {journal} {Phys.
  Rev. C}\ }\textbf {\bibinfo {volume} {84}},\ \bibinfo {pages} {054003}
  (\bibinfo {year} {2011})},\ \Eprint {http://arxiv.org/abs/1107.4956}
  {arXiv:1107.4956 [nucl-th]} \BibitemShut {NoStop}%
\bibitem [{\citenamefont {Neff}\ \emph {et~al.}(2015)\citenamefont {Neff},
  \citenamefont {Feldmeier},\ and\ \citenamefont {Horiuchi}}]{neff15}%
  \BibitemOpen
  \bibfield  {author} {\bibinfo {author} {\bibfnamefont {T.}~\bibnamefont
  {Neff}}, \bibinfo {author} {\bibfnamefont {H.}~\bibnamefont {Feldmeier}}, \
  and\ \bibinfo {author} {\bibfnamefont {W.}~\bibnamefont {Horiuchi}},\ }\href
  {\doibase 10.1103/PhysRevC.92.024003} {\bibfield  {journal} {\bibinfo
  {journal} {Phys. Rev. C}\ }\textbf {\bibinfo {volume} {92}},\ \bibinfo
  {pages} {024003} (\bibinfo {year} {2015})}\BibitemShut {NoStop}%
\bibitem [{\citenamefont {Carlson}\ \emph {et~al.}(2012)\citenamefont
  {Carlson}, \citenamefont {Gandolfi},\ and\ \citenamefont
  {Gezerlis}}]{carlson12}%
  \BibitemOpen
  \bibfield  {author} {\bibinfo {author} {\bibfnamefont {J.}~\bibnamefont
  {Carlson}}, \bibinfo {author} {\bibfnamefont {S.}~\bibnamefont {Gandolfi}}, \
  and\ \bibinfo {author} {\bibfnamefont {A.}~\bibnamefont {Gezerlis}},\
  }\href@noop {} {\bibfield  {journal} {\bibinfo  {journal} {Prog. Theor. Exp.
  Phys.}\ }\textbf {\bibinfo {volume} {01A}},\ \bibinfo {pages} {209} (\bibinfo
  {year} {2012})}\BibitemShut {NoStop}%
\bibitem [{\citenamefont {Wiringa}\ \emph {et~al.}(2014)\citenamefont
  {Wiringa}, \citenamefont {Schiavilla}, \citenamefont {Pieper},\ and\
  \citenamefont {Carlson}}]{wiringa14}%
  \BibitemOpen
  \bibfield  {author} {\bibinfo {author} {\bibfnamefont {R.~B.}\ \bibnamefont
  {Wiringa}}, \bibinfo {author} {\bibfnamefont {R.}~\bibnamefont {Schiavilla}},
  \bibinfo {author} {\bibfnamefont {S.~C.}\ \bibnamefont {Pieper}}, \ and\
  \bibinfo {author} {\bibfnamefont {J.}~\bibnamefont {Carlson}},\ }\href@noop
  {} {\bibfield  {journal} {\bibinfo  {journal} {Phys. Rev. C}\ }\textbf
  {\bibinfo {volume} {89}},\ \bibinfo {pages} {024305} (\bibinfo {year}
  {2014})}\BibitemShut {NoStop}%
\bibitem [{\citenamefont {Lynn}\ \emph {et~al.}(2020)\citenamefont {Lynn},
  \citenamefont {Lonardoni}, \citenamefont {Carlson}, \citenamefont {Chen},
  \citenamefont {Detmold}, \citenamefont {Gandolfi},\ and\ \citenamefont
  {Schwenk}}]{Lynn:2019vwp}%
  \BibitemOpen
  \bibfield  {author} {\bibinfo {author} {\bibfnamefont {J.}~\bibnamefont
  {Lynn}}, \bibinfo {author} {\bibfnamefont {D.}~\bibnamefont {Lonardoni}},
  \bibinfo {author} {\bibfnamefont {J.}~\bibnamefont {Carlson}}, \bibinfo
  {author} {\bibfnamefont {J.}~\bibnamefont {Chen}}, \bibinfo {author}
  {\bibfnamefont {W.}~\bibnamefont {Detmold}}, \bibinfo {author} {\bibfnamefont
  {S.}~\bibnamefont {Gandolfi}}, \ and\ \bibinfo {author} {\bibfnamefont
  {A.}~\bibnamefont {Schwenk}},\ }\href {\doibase 10.1088/1361-6471/ab6af7}
  {\bibfield  {journal} {\bibinfo  {journal} {J. Phys. G}\ }\textbf {\bibinfo
  {volume} {47}},\ \bibinfo {pages} {045109} (\bibinfo {year} {2020})},\
  \Eprint {http://arxiv.org/abs/1903.12587} {arXiv:1903.12587 [nucl-th]}
  \BibitemShut {NoStop}%
\bibitem [{\citenamefont {Pybus}\ \emph {et~al.}(2020)\citenamefont {Pybus},
  \citenamefont {Korover}, \citenamefont {Weiss}, \citenamefont {Schmidt},
  \citenamefont {Barnea}, \citenamefont {Higinbotham}, \citenamefont
  {Piasetzky}, \citenamefont {Strikman}, \citenamefont {Weinstein},\ and\
  \citenamefont {Hen}}]{Pybus:2020itv}%
  \BibitemOpen
  \bibfield  {author} {\bibinfo {author} {\bibfnamefont {J.}~\bibnamefont
  {Pybus}}, \bibinfo {author} {\bibfnamefont {I.}~\bibnamefont {Korover}},
  \bibinfo {author} {\bibfnamefont {R.}~\bibnamefont {Weiss}}, \bibinfo
  {author} {\bibfnamefont {A.}~\bibnamefont {Schmidt}}, \bibinfo {author}
  {\bibfnamefont {N.}~\bibnamefont {Barnea}}, \bibinfo {author} {\bibfnamefont
  {D.}~\bibnamefont {Higinbotham}}, \bibinfo {author} {\bibfnamefont
  {E.}~\bibnamefont {Piasetzky}}, \bibinfo {author} {\bibfnamefont
  {M.}~\bibnamefont {Strikman}}, \bibinfo {author} {\bibfnamefont
  {L.}~\bibnamefont {Weinstein}}, \ and\ \bibinfo {author} {\bibfnamefont
  {O.}~\bibnamefont {Hen}},\ }\href {\doibase 10.1016/j.physletb.2020.135429}
  {\bibfield  {journal} {\bibinfo  {journal} {Phys. Lett. B}\ }\textbf
  {\bibinfo {volume} {805}},\ \bibinfo {pages} {135429} (\bibinfo {year}
  {2020})},\ \Eprint {http://arxiv.org/abs/2003.02318} {arXiv:2003.02318
  [nucl-th]} \BibitemShut {NoStop}%
\bibitem [{\citenamefont {Colle}\ \emph {et~al.}(2014)\citenamefont {Colle},
  \citenamefont {Cosyn}, \citenamefont {Ryckebusch},\ and\ \citenamefont
  {Vanhalst}}]{Colle:2013nna}%
  \BibitemOpen
  \bibfield  {author} {\bibinfo {author} {\bibfnamefont {C.}~\bibnamefont
  {Colle}}, \bibinfo {author} {\bibfnamefont {W.}~\bibnamefont {Cosyn}},
  \bibinfo {author} {\bibfnamefont {J.}~\bibnamefont {Ryckebusch}}, \ and\
  \bibinfo {author} {\bibfnamefont {M.}~\bibnamefont {Vanhalst}},\ }\href
  {\doibase 10.1103/PhysRevC.89.024603} {\bibfield  {journal} {\bibinfo
  {journal} {Phys. Rev.}\ }\textbf {\bibinfo {volume} {C89}},\ \bibinfo {pages}
  {024603} (\bibinfo {year} {2014})}\BibitemShut {NoStop}%
\bibitem [{\citenamefont {Wiringa}\ \emph {et~al.}(1995)\citenamefont
  {Wiringa}, \citenamefont {Stoks},\ and\ \citenamefont
  {Schiavilla}}]{wiringa95}%
  \BibitemOpen
  \bibfield  {author} {\bibinfo {author} {\bibfnamefont {R.~B.}\ \bibnamefont
  {Wiringa}}, \bibinfo {author} {\bibfnamefont {V.~G.~J.}\ \bibnamefont
  {Stoks}}, \ and\ \bibinfo {author} {\bibfnamefont {R.}~\bibnamefont
  {Schiavilla}},\ }\href@noop {} {\bibfield  {journal} {\bibinfo  {journal}
  {Phys. Rev. C}\ }\textbf {\bibinfo {volume} {51}},\ \bibinfo {pages} {38}
  (\bibinfo {year} {1995})}\BibitemShut {NoStop}%
\bibitem [{\citenamefont {Ryckebusch}\ \emph {et~al.}(2019)\citenamefont
  {Ryckebusch}, \citenamefont {Cosyn}, \citenamefont {Vieijra},\ and\
  \citenamefont {Casert}}]{Ryckebusch:2019oya}%
  \BibitemOpen
  \bibfield  {author} {\bibinfo {author} {\bibfnamefont {J.}~\bibnamefont
  {Ryckebusch}}, \bibinfo {author} {\bibfnamefont {W.}~\bibnamefont {Cosyn}},
  \bibinfo {author} {\bibfnamefont {T.}~\bibnamefont {Vieijra}}, \ and\
  \bibinfo {author} {\bibfnamefont {C.}~\bibnamefont {Casert}},\ }\href
  {\doibase 10.1103/PhysRevC.100.054620} {\bibfield  {journal} {\bibinfo
  {journal} {Phys. Rev. C}\ }\textbf {\bibinfo {volume} {100}},\ \bibinfo
  {pages} {054620} (\bibinfo {year} {2019})},\ \Eprint
  {http://arxiv.org/abs/1907.07259} {arXiv:1907.07259 [nucl-th]} \BibitemShut
  {NoStop}%
\bibitem [{\citenamefont {Wiringa}\ and\ \citenamefont
  {Pieper}(2002)}]{Wiringa:2002}%
  \BibitemOpen
  \bibfield  {author} {\bibinfo {author} {\bibfnamefont {R.~B.}\ \bibnamefont
  {Wiringa}}\ and\ \bibinfo {author} {\bibfnamefont {S.~C.}\ \bibnamefont
  {Pieper}},\ }\href {\doibase 10.1103/PhysRevLett.89.182501} {\bibfield
  {journal} {\bibinfo  {journal} {Phys. Rev. Lett.}\ }\textbf {\bibinfo
  {volume} {89}},\ \bibinfo {pages} {182501} (\bibinfo {year}
  {2002})}\BibitemShut {NoStop}%
\bibitem [{\citenamefont {Piarulli}\ \emph {et~al.}(2016)\citenamefont
  {Piarulli}, \citenamefont {Girlanda}, \citenamefont {Schiavilla},
  \citenamefont {Kievsky}, \citenamefont {Lovato}, \citenamefont {Marcucci},
  \citenamefont {Pieper}, \citenamefont {Viviani},\ and\ \citenamefont
  {Wiringa}}]{Piarulli:2016vel}%
  \BibitemOpen
  \bibfield  {author} {\bibinfo {author} {\bibfnamefont {M.}~\bibnamefont
  {Piarulli}}, \bibinfo {author} {\bibfnamefont {L.}~\bibnamefont {Girlanda}},
  \bibinfo {author} {\bibfnamefont {R.}~\bibnamefont {Schiavilla}}, \bibinfo
  {author} {\bibfnamefont {A.}~\bibnamefont {Kievsky}}, \bibinfo {author}
  {\bibfnamefont {A.}~\bibnamefont {Lovato}}, \bibinfo {author} {\bibfnamefont
  {L.~E.}\ \bibnamefont {Marcucci}}, \bibinfo {author} {\bibfnamefont {S.~C.}\
  \bibnamefont {Pieper}}, \bibinfo {author} {\bibfnamefont {M.}~\bibnamefont
  {Viviani}}, \ and\ \bibinfo {author} {\bibfnamefont {R.~B.}\ \bibnamefont
  {Wiringa}},\ }\href {\doibase 10.1103/PhysRevC.94.054007} {\bibfield
  {journal} {\bibinfo  {journal} {Phys. Rev.}\ }\textbf {\bibinfo {volume}
  {C94}},\ \bibinfo {pages} {054007} (\bibinfo {year} {2016})}\BibitemShut
  {NoStop}%
\bibitem [{\citenamefont {Piarulli}\ \emph {et~al.}(2018)\citenamefont
  {Piarulli} \emph {et~al.}}]{Piarulli:2017dwd}%
  \BibitemOpen
  \bibfield  {author} {\bibinfo {author} {\bibfnamefont {M.}~\bibnamefont
  {Piarulli}} \emph {et~al.},\ }\href {\doibase 10.1103/PhysRevLett.120.052503}
  {\bibfield  {journal} {\bibinfo  {journal} {Phys. Rev. Lett.}\ }\textbf
  {\bibinfo {volume} {120}},\ \bibinfo {pages} {052503} (\bibinfo {year}
  {2018})}\BibitemShut {NoStop}%
\bibitem [{\citenamefont {Baroni}\ \emph {et~al.}(2018)\citenamefont {Baroni},
  \citenamefont {Schiavilla}, \citenamefont {Marcucci}, \citenamefont
  {Girlanda}, \citenamefont {Kievsky}, \citenamefont {Lovato}, \citenamefont
  {Pastore}, \citenamefont {Piarulli}, \citenamefont {Pieper}, \citenamefont
  {Viviani},\ and\ \citenamefont {Wiringa}}]{Baroni:2018fdn}%
  \BibitemOpen
  \bibfield  {author} {\bibinfo {author} {\bibfnamefont {A.}~\bibnamefont
  {Baroni}}, \bibinfo {author} {\bibfnamefont {R.}~\bibnamefont {Schiavilla}},
  \bibinfo {author} {\bibfnamefont {L.~E.}\ \bibnamefont {Marcucci}}, \bibinfo
  {author} {\bibfnamefont {L.}~\bibnamefont {Girlanda}}, \bibinfo {author}
  {\bibfnamefont {A.}~\bibnamefont {Kievsky}}, \bibinfo {author} {\bibfnamefont
  {A.}~\bibnamefont {Lovato}}, \bibinfo {author} {\bibfnamefont
  {S.}~\bibnamefont {Pastore}}, \bibinfo {author} {\bibfnamefont
  {M.}~\bibnamefont {Piarulli}}, \bibinfo {author} {\bibfnamefont {S.~C.}\
  \bibnamefont {Pieper}}, \bibinfo {author} {\bibfnamefont {M.}~\bibnamefont
  {Viviani}}, \ and\ \bibinfo {author} {\bibfnamefont {R.~B.}\ \bibnamefont
  {Wiringa}},\ }\href {\doibase 10.1103/PhysRevC.98.044003} {\bibfield
  {journal} {\bibinfo  {journal} {Phys. Rev. C}\ }\textbf {\bibinfo {volume}
  {98}},\ \bibinfo {pages} {044003} (\bibinfo {year} {2018})}\BibitemShut
  {NoStop}%
\bibitem [{\citenamefont {Gezerlis}\ \emph {et~al.}(2014)\citenamefont
  {Gezerlis}, \citenamefont {Tews}, \citenamefont {Epelbaum}, \citenamefont
  {Freunek}, \citenamefont {Gandolfi}, \citenamefont {Hebeler}, \citenamefont
  {Nogga},\ and\ \citenamefont {Schwenk}}]{Gezerlis:2014}%
  \BibitemOpen
  \bibfield  {author} {\bibinfo {author} {\bibfnamefont {A.}~\bibnamefont
  {Gezerlis}}, \bibinfo {author} {\bibfnamefont {I.}~\bibnamefont {Tews}},
  \bibinfo {author} {\bibfnamefont {E.}~\bibnamefont {Epelbaum}}, \bibinfo
  {author} {\bibfnamefont {M.}~\bibnamefont {Freunek}}, \bibinfo {author}
  {\bibfnamefont {S.}~\bibnamefont {Gandolfi}}, \bibinfo {author}
  {\bibfnamefont {K.}~\bibnamefont {Hebeler}}, \bibinfo {author} {\bibfnamefont
  {A.}~\bibnamefont {Nogga}}, \ and\ \bibinfo {author} {\bibfnamefont
  {A.}~\bibnamefont {Schwenk}},\ }\href {\doibase 10.1103/PhysRevC.90.054323}
  {\bibfield  {journal} {\bibinfo  {journal} {Phys. Rev. C}\ }\textbf {\bibinfo
  {volume} {90}},\ \bibinfo {pages} {054323} (\bibinfo {year}
  {2014})}\BibitemShut {NoStop}%
\bibitem [{\citenamefont {Lynn}\ \emph {et~al.}(2016)\citenamefont {Lynn},
  \citenamefont {Tews}, \citenamefont {Carlson}, \citenamefont {Gandolfi},
  \citenamefont {Gezerlis}, \citenamefont {Schmidt},\ and\ \citenamefont
  {Schwenk}}]{Lynn:2016}%
  \BibitemOpen
  \bibfield  {author} {\bibinfo {author} {\bibfnamefont {J.~E.}\ \bibnamefont
  {Lynn}}, \bibinfo {author} {\bibfnamefont {I.}~\bibnamefont {Tews}}, \bibinfo
  {author} {\bibfnamefont {J.}~\bibnamefont {Carlson}}, \bibinfo {author}
  {\bibfnamefont {S.}~\bibnamefont {Gandolfi}}, \bibinfo {author}
  {\bibfnamefont {A.}~\bibnamefont {Gezerlis}}, \bibinfo {author}
  {\bibfnamefont {K.~E.}\ \bibnamefont {Schmidt}}, \ and\ \bibinfo {author}
  {\bibfnamefont {A.}~\bibnamefont {Schwenk}},\ }\href {\doibase
  10.1103/PhysRevLett.116.062501} {\bibfield  {journal} {\bibinfo  {journal}
  {Phys. Rev. Lett.}\ }\textbf {\bibinfo {volume} {116}},\ \bibinfo {pages}
  {062501} (\bibinfo {year} {2016})}\BibitemShut {NoStop}%
\bibitem [{\citenamefont {Lonardoni}\ \emph {et~al.}(2018)\citenamefont
  {Lonardoni}, \citenamefont {Gandolfi}, \citenamefont {Lynn}, \citenamefont
  {Petrie}, \citenamefont {Carlson}, \citenamefont {Schmidt},\ and\
  \citenamefont {Schwenk}}]{Lonardoni:2018prc}%
  \BibitemOpen
  \bibfield  {author} {\bibinfo {author} {\bibfnamefont {D.}~\bibnamefont
  {Lonardoni}}, \bibinfo {author} {\bibfnamefont {S.}~\bibnamefont {Gandolfi}},
  \bibinfo {author} {\bibfnamefont {J.~E.}\ \bibnamefont {Lynn}}, \bibinfo
  {author} {\bibfnamefont {C.}~\bibnamefont {Petrie}}, \bibinfo {author}
  {\bibfnamefont {J.}~\bibnamefont {Carlson}}, \bibinfo {author} {\bibfnamefont
  {K.~E.}\ \bibnamefont {Schmidt}}, \ and\ \bibinfo {author} {\bibfnamefont
  {A.}~\bibnamefont {Schwenk}},\ }\href {\doibase 10.1103/PhysRevC.97.044318}
  {\bibfield  {journal} {\bibinfo  {journal} {Phys. Rev. C}\ }\textbf {\bibinfo
  {volume} {97}},\ \bibinfo {pages} {044318} (\bibinfo {year}
  {2018})}\BibitemShut {NoStop}%
\bibitem [{\citenamefont {Vanhalst}\ \emph {et~al.}(2012)\citenamefont
  {Vanhalst}, \citenamefont {Ryckebusch},\ and\ \citenamefont
  {Cosyn}}]{vanhalst12}%
  \BibitemOpen
  \bibfield  {author} {\bibinfo {author} {\bibfnamefont {M.}~\bibnamefont
  {Vanhalst}}, \bibinfo {author} {\bibfnamefont {J.}~\bibnamefont
  {Ryckebusch}}, \ and\ \bibinfo {author} {\bibfnamefont {W.}~\bibnamefont
  {Cosyn}},\ }\href {\doibase 10.1103/PhysRevC.86.044619} {\bibfield  {journal}
  {\bibinfo  {journal} {Phys. Rev. C}\ }\textbf {\bibinfo {volume} {86}},\
  \bibinfo {pages} {044619} (\bibinfo {year} {2012})}\BibitemShut {NoStop}%
\bibitem [{\citenamefont {Arrington}\ \emph {et~al.}(2012)\citenamefont
  {Arrington}, \citenamefont {Daniel}, \citenamefont {Day}, \citenamefont
  {Fomin}, \citenamefont {Gaskell},\ and\ \citenamefont
  {Solvignon}}]{Arrington12}%
  \BibitemOpen
  \bibfield  {author} {\bibinfo {author} {\bibfnamefont {J.}~\bibnamefont
  {Arrington}}, \bibinfo {author} {\bibfnamefont {A.}~\bibnamefont {Daniel}},
  \bibinfo {author} {\bibfnamefont {D.~B.}\ \bibnamefont {Day}}, \bibinfo
  {author} {\bibfnamefont {N.}~\bibnamefont {Fomin}}, \bibinfo {author}
  {\bibfnamefont {D.}~\bibnamefont {Gaskell}}, \ and\ \bibinfo {author}
  {\bibfnamefont {P.}~\bibnamefont {Solvignon}},\ }\href {\doibase
  10.1103/PhysRevC.86.065204} {\bibfield  {journal} {\bibinfo  {journal} {Phys.
  Rev. C}\ }\textbf {\bibinfo {volume} {86}},\ \bibinfo {pages} {065204}
  (\bibinfo {year} {2012})}\BibitemShut {NoStop}%
\bibitem [{\citenamefont {Sargsian}\ \emph {et~al.}(2019)\citenamefont
  {Sargsian}, \citenamefont {Day}, \citenamefont {Frankfurt},\ and\
  \citenamefont {Strikman}}]{Sargsian:2019joj}%
  \BibitemOpen
  \bibfield  {author} {\bibinfo {author} {\bibfnamefont {M.~M.}\ \bibnamefont
  {Sargsian}}, \bibinfo {author} {\bibfnamefont {D.~B.}\ \bibnamefont {Day}},
  \bibinfo {author} {\bibfnamefont {L.~L.}\ \bibnamefont {Frankfurt}}, \ and\
  \bibinfo {author} {\bibfnamefont {M.~I.}\ \bibnamefont {Strikman}},\ }\href
  {\doibase 10.1103/PhysRevC.100.044320} {\bibfield  {journal} {\bibinfo
  {journal} {Phys. Rev. C}\ }\textbf {\bibinfo {volume} {100}},\ \bibinfo
  {pages} {044320} (\bibinfo {year} {2019})},\ \Eprint
  {http://arxiv.org/abs/1910.14663} {arXiv:1910.14663 [nucl-th]} \BibitemShut
  {NoStop}%
\bibitem [{\citenamefont {Sargsian}(2014)}]{sargsian14}%
  \BibitemOpen
  \bibfield  {author} {\bibinfo {author} {\bibfnamefont {M.~M.}\ \bibnamefont
  {Sargsian}},\ }\href {\doibase 10.1103/PhysRevC.89.034305} {\bibfield
  {journal} {\bibinfo  {journal} {Phys. Rev. C}\ }\textbf {\bibinfo {volume}
  {89}},\ \bibinfo {pages} {034305} (\bibinfo {year} {2014})}\BibitemShut
  {NoStop}%
\bibitem [{\citenamefont {McGauley}\ and\ \citenamefont
  {Sargsian}(2011)}]{McGauley:2011qc}%
  \BibitemOpen
  \bibfield  {author} {\bibinfo {author} {\bibfnamefont {M.}~\bibnamefont
  {McGauley}}\ and\ \bibinfo {author} {\bibfnamefont {M.~M.}\ \bibnamefont
  {Sargsian}},\ }\href@noop {} {\  (\bibinfo {year} {2011})},\ \Eprint
  {http://arxiv.org/abs/1102.3973} {arXiv:1102.3973 [nucl-th]} \BibitemShut
  {NoStop}%
\bibitem [{\citenamefont {More}\ \emph {et~al.}(2017)\citenamefont {More},
  \citenamefont {Bogner},\ and\ \citenamefont {Furnstahl}}]{More:2017syr}%
  \BibitemOpen
  \bibfield  {author} {\bibinfo {author} {\bibfnamefont {S.~N.}\ \bibnamefont
  {More}}, \bibinfo {author} {\bibfnamefont {S.~K.}\ \bibnamefont {Bogner}}, \
  and\ \bibinfo {author} {\bibfnamefont {R.~J.}\ \bibnamefont {Furnstahl}},\
  }\href {\doibase 10.1103/PhysRevC.96.054004} {\bibfield  {journal} {\bibinfo
  {journal} {Phys.\ Rev.\ C}\ }\textbf {\bibinfo {volume} {96}},\ \bibinfo
  {pages} {054004} (\bibinfo {year} {2017})},\ \Eprint
  {http://arxiv.org/abs/1708.03315} {arXiv:1708.03315 [nucl-th]} \BibitemShut
  {NoStop}%
\end{thebibliography}%

\end{document}